\documentclass[11pt]{article}
\pdfoutput=1  % togliere il commento se le figure sono pdf, lasciarlo se sono eps
\usepackage{graphicx,color}
\usepackage{appendix}
\usepackage{latexsym,amsmath,amssymb,graphicx,booktabs}
\usepackage{epsfig,latexsym,cite}
\usepackage{hyperref}
\numberwithin{equation}{section}% numera le equazioni seconde le sezioni , e.g. 1.15 invece che consecutivamente; anche le appendici, eq. (A.1) etc. Richiede amsmath

\definecolor{MyBlue}{rgb}{0.15,0.15,0.70}

\hypersetup{
colorlinks=true,
citecolor=MyBlue,
linkcolor=MyBlue,
urlcolor=MyBlue
}

\input epsf
\usepackage{amssymb}
\usepackage{amsmath}
\usepackage{amsfonts}
\usepackage{upgreek}
\usepackage{latexsym}

\newcommand{\iBox}{\Box^{-1}}
\renewcommand\({\left(}
\renewcommand\){\right)}
\renewcommand\[{\left[}
\renewcommand\]{\right]}

\newcommand\n{{\mbox {\boldmath $\nabla$}}}
\newcommand{\ra}{\rightarrow}

\def\lsim{\raise 0.4ex\hbox{$<$}\kern -0.8em\lower 0.62
ex\hbox{$\sim$}}

\def\gsim{\raise 0.4ex\hbox{$>$}\kern -0.7em\lower 0.62
ex\hbox{$\sim$}}

\def\lbar{{\hbox{$\lambda$}\kern -0.7em\raise 0.6ex
\hbox{$-$}}}

\newcommand\eq[1]{eq.~(\ref{#1})}
\newcommand\eqs[2]{eqs.~(\ref{#1}) and (\ref{#2})}
\newcommand\Eq[1]{Equation~(\ref{#1})}
\newcommand\Eqs[2]{Equations~(\ref{#1}) and (\ref{#2})}

\newcommand\eqss[3]{eqs.~(\ref{#1}), (\ref{#2}) and (\ref{#3})}
\newcommand\eqsss[4]{eqs.~(\ref{#1}), (\ref{#2}), (\ref{#3})
and (\ref{#4})}

\newcommand\eqst[2]{eqs.~(\ref{#1})--(\ref{#2})}

\newcommand\pa{\partial}
\newcommand\p{\partial}

\newcommand\ee{\end{equation}}
\newcommand\be{\begin{equation}}
\def\bea{\begin{array}}
\def\eea{\end{array}}\def\ea{\end{array}}
\newcommand\ees{\end{eqnarray}}
\newcommand\bees{\begin{eqnarray}}
\def\nn{\nonumber}

\def\a{\alpha}
\def\b{\beta}

\def\s{\sigma}

\def\d{\delta}

\def\eps{\epsilon}

\def\dslash{\hspace{-1mm}\not{\hbox{\kern-2pt $\partial$}}}
\def\Dslash{\not{\hbox{\kern-4pt $D$}}}
\def\pslash{\not{\hbox{\kern-2.1pt $p$}}}
\def\kslash{\not{\hbox{\kern-2.3pt $k$}}}
\def\qslash{\not{\hbox{\kern-2.3pt $q$}}}

\newcommand{\vk}{{\bf k}}
\newcommand{\vx}{{\bf x}}

\def\p1{{\bf p}_1}
\def\p2{{\bf p}_2}
\def\k1{{\bf k}_1}
\def\k2{{\bf k}_2}

\newcommand{\emn}{\eta_{\mu\nu}}

\newcommand{\eMN}{\eta^{\mu\nu}}
\newcommand{\eRS}{\eta^{\rho\sigma}}
\newcommand{\eMR}{\eta^{\mu\rho}}
\newcommand{\eNS}{\eta^{\nu\sigma}}
\newcommand{\eMS}{\eta^{\mu\sigma}}
\newcommand{\eNR}{\eta^{\nu\rho}}

\newcommand{\gmn}{g_{\mu\nu}}

\newcommand{\gMN}{g^{\mu\nu}}

\newcommand{\gbmn}{\bar{g}_{\mu\nu}}

\newcommand{\hmn}{h_{\mu\nu}}
\newcommand{\hrs}{h_{\rho\sigma}}

\newcommand{\hMN}{h^{\mu\nu}}

\newcommand{\hatn}{\hat{\bf n}}

\newcommand{\xim}{\xi_{\mu}}
\newcommand{\xin}{\xi_{\nu}}

\newcommand{\pam}{\pa_{\mu}}

\newcommand{\pan}{\pa_{\nu}}
\newcommand{\parho}{\pa_{\rho}}
\newcommand{\pas}{\pa_{\sigma}}
\newcommand{\paM}{\pa^{\mu}}
\newcommand{\paN}{\pa^{\nu}}

\newcommand{\Rmn}{R_{\mu\nu}}
\newcommand{\Gmn}{G_{\mu\nu}}

\newcommand{\Rmnrs}{R_{\mu\nu\rho\sigma}}

\newcommand{\Tmn}{T_{\mu\nu}}
\newcommand{\Smn}{S_{\mu\nu}}

\newcommand{\TMN}{T^{\mu\nu}}

\newcommand{\dddM}{\kern 0.2em \raise 1.9ex\hbox{$...$}\kern -1.0em \hbox{$M$}}
\newcommand{\dddQ}{\kern 0.2em \raise 1.9ex\hbox{$...$}\kern -1.0em \hbox{$Q$}}
\newcommand{\dddI}{\kern 0.2em \raise 1.9ex\hbox{$...$}\kern -1.0em\hbox{$I$}}
\newcommand{\dddJ}{\kern 0.2em \raise 1.9ex\hbox{$...$}\kern-1.0em
\hbox{$J$}}
\newcommand{\dddcalJ}{\kern 0.2em \raise 1.9ex\hbox{$...$}\kern-1.0em
\hbox{${\cal J}$}}

\newcommand{\dddO}{\kern 0.2em \raise 1.9ex\hbox{$...$}\kern -1.0em
\hbox{${\cal O}$}}
\def\dddz{\raise 1.5ex\hbox{$...$}\kern -0.8em \hbox{$z$}}
\def\dddd{\raise 1.8ex\hbox{$...$}\kern -0.8em \hbox{$d$}}
\def\dddbd{\raise 1.8ex\hbox{$...$}\kern -0.8em \hbox{${\bf d}$}}
\def\ddbd{\raise 1.8ex\hbox{$..$}\kern -0.8em \hbox{${\bf d}$}}
\def\dddx{\raise 1.6ex\hbox{$...$}\kern -0.8em \hbox{$x$}}

\newcommand{\msun}{M_{\odot}}

\newcommand{\Sch}{Schwarzschild }

\newcommand{\mpl}{M_{\rm Pl}}

\newcommand{\ode}{\Omega_{\rm DE}}

\begin{document}

\begin{titlepage}

\vspace*{2cm}

\centerline{\Large \bf Spherically symmetric static solutions in a nonlocal}

\vspace{5mm}

\centerline{\Large \bf infrared modification of General Relativity}

\vskip 0.4cm
\vskip 0.7cm
\centerline{\large  Alex Kehagias$^{a,b}$ and  Michele Maggiore$^a$}
\vskip 0.3cm
\centerline{\em $^a$D\'epartement de Physique Th\'eorique and Center for Astroparticle Physics,}  
\centerline{\em Universit\'e de Gen\`eve, 24 quai Ansermet, CH--1211 Gen\`eve 4, Switzerland}
\vspace{3mm}
\centerline{\em $^b$Physics Division, National Technical University of Athens,\\}  
\centerline{\em 15780 Zografou Campus, Athens, Greece}

\vskip 1.9cm

\begin{abstract}

We discuss static spherically symmetric solutions in a recently proposed  nonlocal infrared modification of Einstein equations induced by  a term $m^2g_{\mu\nu}\Box^{-1} R$, where $m$ is a mass scale. We find that, 
contrary to what happens in usual theories of massive gravity, in
this nonlocal theory there is no vDVZ discontinuity and classical non-linearities do not become large below a Vainshtein radius  parametrically larger than the Schwarzschild radius $r_S$. Rather on the contrary, in the regime
$r\ll m^{-1}$
the corrections to the metric generated by a static body in GR are of the form $1+{\cal O}(m^2r^2)$ and become smaller and smaller toward smaller values of $r$. The modification to the GR solutions only show up at $r\,\gsim\, m^{-1}$. For $m={\cal O}(H_0)$, as required for having interesting cosmological consequences, the  nonlocal theory therefore recovers all successes of GR at the solar system and lab scales.

\end{abstract}

%\vspace{3cm}

\end{titlepage}

\newpage

\section{Introduction}

Recent years have witnessed  very intense activity on
the study of infrared modifications of General Relativity (GR). This is motivated  by the aim of explaining the present phase of accelerated expansion of the Universe through a modification of gravity at distances $r\sim H_0^{-1}$ (where $H_0$ is the present value of the Hubble parameter), and also turns out to be a rich and challenging theoretical subject. To modify GR in the far infrared it is natural to introduce a mass parameter $m\sim H_0$.  At first sight, this could be achieved by giving a mass to the graviton. However, constructing a consistent theory of massive gravity turns out to be remarkably difficult, and the problem has a long history that goes back  to  classic papers by
Fierz and Pauli  in 1939 \cite{Fierz:1939ix}
and Boulware and Deser in 1973 \cite{Boulware:1973my}. In recent years  there has been significant progress in this direction, in particular with the construction of the ghost-free dRGT  theory 
\cite{deRham:2010ik,deRham:2010kj} (see also \cite{deRham:2011rn,Hassan:2011hr,Hassan:2011tf,Hassan:2011ea}, and \cite{Hinterbichler:2011tt,deRham:2014zqa} for  reviews), although a number of difficulties and open problems persist; the dRGT theory (as well as galileon theories) very likely admits superluminal excitations 
over some backgrounds \cite{Dubovsky:2005xd,Nicolis:2009qm,Gruzinov:2011sq,deRham:2011pt,Deser:2012qx,Berezhiani:2013dw,Izumi:2013poa,Berezhiani:2013dca,Koyama:2013paa}. Furthermore,
even if the sixth ghost-like degree of freedom is absent in any background, the fluctuations of the remaining five degrees of freedom can become ghost-like over non-trivial backgrounds \cite{Gumrukcuoglu:2011ew,Gumrukcuoglu:2011zh,Koyama:2011wx}. Another open problem of dRGT theory is that it is not clear whether a  satisfying cosmology emerges. Homogeneous and isotropic spatially flat  Friedman-Robertson-Walker (FRW) solutions do not exist, and are in fact forbidden by  the same constraint that removes the ghost \cite{D'Amico:2011jj}. There are open isotropic FRW solutions, which however suffer of  strong coupling and ghost-like instabilities \cite{DeFelice:2013awa}.  It is presently unclear whether there are stable  and observationally viable inhomogeneous solutions in the full non-linear theory away from the decoupling 
limit (see the discussion in~\cite{deRham:2014zqa}).

A peculiar  aspect of massive gravity theories 
 is that they require the introduction of an external reference metric. In  the recent papers \cite{Jaccard:2013gla,Maggiore:2013mea} it has been proposed a different approach,  in which gravity is deformed by the introduction of a mass parameter $m$ in such a way that no external reference metric is introduced, and general covariance is preserved. This can be achieved by adding nonlocal terms to the Einstein equations. The introduction of nonlocal terms for producing IR modifications of gravity has been suggested by various authors, following different lines of reasoning. In particular, nonlocal operators that modify GR in the  IR appear in the degravitation proposal \cite{ArkaniHamed:2002fu,Dvali:2006su} (see also \cite{Dvali:2000xg,Dvali:2002pe}). Non-local covariantizations of the Fierz-Pauli theory were discussed in \cite{Porrati:2002cp}.
A different nonlocal cosmological model  has been  proposed in 
\cite{Deser:2007jk}, and  has been further studied in a number of recent papers\cite{Koivisto:2008xfa,Koivisto:2008dh,Capozziello:2008gu,Elizalde:2011su,Zhang:2011uv,Elizalde:2012ja,Park:2012cp,Bamba:2012ky,Deser:2013uya,Ferreira:2013tqn,Dodelson:2013sma} (see \cite{Woodard:2014iga} for a recent review). Another interesting nonlocal model  has been studied in 
\cite{Barvinsky:2003kg,Barvinsky:2011hd,Barvinsky:2011rk}. 
The model that has been proposed by one of us in  \cite{Maggiore:2013mea} is defined by the classical equation of motion
\be\label{modela2}
\Gmn -\frac{m^2}{3}\, \(\gmn \iBox_{\rm ret} R\)^{\rm T}=8\pi G\,\Tmn\, .
\ee
The inverse of the d'Alembertian is defined with the retarded Green's function, which ensures causality.
The superscript T denotes the extraction of the transverse part of the tensor and ensures that the left-hand side
of \eq{modela2} has zero divergence, and therefore $\Tmn$ is automatically conserved. The extraction of the transverse part exploits the fact that, in a generic curved space-time, any symmetric tensor $\Smn$ can be decomposed as
\be\label{splitSmn}
S_{\mu\nu}=S_{\mu\nu}^{\rm T}+\frac{1}{2}(\n_{\mu}S_{\nu}+\n_{\nu}S_{\mu})\, , 
\ee
where
$\n^{\mu}S_{\mu\nu}^{\rm T}=0$ \cite{Deser:1967zzb,York:1974}.  The factor $1/3$ 
in \eq{modela2} is a convenient normalization of the parameter $m^2$ in $d=3$ spatial dimensions (and becomes $(d-1)/(2d)$ for generic $d$). Some conceptual aspects of this model have been discussed in \cite{Maggiore:2013mea,Foffa:2013sma}. Its cosmological consequences, at the level of background evolution, have been studied in
\cite{Maggiore:2013mea,Foffa:2013vma}, while a study of its cosmological perturbations will be presented in \cite{Dirian:2014ara}.

At the conceptual level, it is important to stress that the $\iBox$ operator
in \eq{modela2} is defined with the retarded Green's function. This ensures causality, and also has the important consequence that \eq{modela2} cannot be  the equation of motion of a fundamental nonlocal QFT. Indeed, the variation of an action involving $\iBox$ always gives rise to an equation of motion involving a symmetrized Green's function, rather than a retarded one \cite{Deser:2007jk,Barvinsky:2011rk,Jaccard:2013gla}. \Eq{modela2} should rather be understood  as a classical effective equation of motion. Non-local effective equations involving a retarded Green's function
govern for instance the  dynamics of the in-in matrix elements of quantum fields, such as 
$\langle 0_{\rm in}|\hat{\phi}|0_{\rm in}\rangle$ or $\langle 0_{\rm in}|\hat{g}_{\mu\nu}|0_{\rm in}\rangle$, and encode  quantum corrections  
to the classical dynamics \cite{Jordan:1986ug,Calzetta:1986ey}. 
Thus, issues of quantum vacuum decay induced by ghost instabilities, or non-linearities induced by quantum corrections, cannot be addressed directly from a study of
\eq{modela2}, but should rather be addressed in the fundamental underlying (local) QFT. This, however, is not conceptually very different from what happens in dRGT, where the UV completion is needed to address the causality issue. 

Observe that one can still in principle derive non-local causal equations from an action, using the   formal trick  
of replacing by hand
$\iBox\ra\iBox_{\rm ret}$ after performing the variation. This is  indeed the 
procedure used in \cite{Soussa:2003vv,Deser:2007jk}, in the context of non-local  gravity theories with a Lagrangian of the form $Rf(\iBox R)$, see also the  recent discussion in \cite{Tsamis:2014hra}.
However, any direct connection to a fundamental {\em quantum} field theory  is then lost. One can ask what is the classical non-local action that, in the above sense, reproduce \eq{modela2}. If one linearizes \eq{modela2} over flat Minkowski space, writing $\gmn=\emn+\hmn$,  one gets~\cite{Maggiore:2013mea}
\be\label{line1}
{\cal E}^{\mu\nu,\rho\sigma}\hrs
-\frac{2}{3}\, m^2 P^{\mu\nu}P^{\rho\sigma}
\hrs +{\cal O}(h^2)=-16\pi G\TMN\, ,
\ee
where  ${\cal E}^{\mu\nu,\rho\sigma}$  is the  Lichnerowicz operator,
\be
P^{\mu\nu}=\eMN-\frac{\paM\paN}{\Box}\, ,
\ee
and $\Box$ is now  the   flat-space d'Alembertian. This linearized equation can of course be derived from the action
\be
S=\int d^4x \[ \frac{1}{4}\hmn{\cal E}^{\mu\nu,\rho\sigma}\hrs
-\frac{1}{6}\, m^2 \hmn P^{\mu\nu}P^{\rho\sigma}
\hrs +{\cal O}(h^3)\] +S_M\, ,
\ee
where $S_M$ is the matter action. This can be rewritten in a covariant form, as
\be\label{S1R3}
S=\frac{1}{16\pi G}\int d^{4}x \sqrt{-g}\, 
\[R-\frac{1}{6} m^2R\frac{1}{\Box^2} R +{\cal O}(\Rmnrs^3)\] +S_M\, .
\ee
Thus, the  action corresponding to \eq{modela2} contains a full series of cubic and higher-order terms in the curvature, and we do not have a compact closed-form expression. The model obtained truncating \eq{S1R3} to order $R^2$ is interesting in its own right, and has been studied in \cite{Maggiore:2014sia}. Observe that the cubic and higher order terms are suppressed by powers of $\iBox R={\cal O}(h)$ (and not of $R/\mpl^2$, where $\mpl$ is the Planck mass). They are therefore on the same footing as the usual non-linearities of GR and cannot be neglected on non-trivial backgrounds, e.g. in cosmological backgrounds. So, the model defined by
\be\label{SNLnoR3}
S\equiv \frac{1}{16\pi G}\int d^{4}x \sqrt{-g}\, 
\[R-\frac{1}{6} m^2R\frac{1}{\Box^2} R \] +S_M\, ,
\ee
with no cubic and higher-order terms, 
and the model defined by \eq{modela2} are different. In a sense, \eq{modela2} provides the simplest non-local equation of motion in this class of model involving $\iBox R$ and a mass scale $m$, while \eq{SNLnoR3} provides the simplest action.

At the phenomenological level, \eq{modela2} turns out to have rather interesting consequences. In particular, at the level of background evolution it admits flat FRW solutions, in which furthermore a dynamical dark energy emerges automatically. By fixing the free parameter $m$ to a value $m\simeq 0.67 H_0$ the model reproduces the observed value $\ode\simeq 0.68$. This leaves us with no free parameter and we then get a pure prediction for the EOS parameter of dark energy. Using the standard fit of the form $w_{\rm DE}(a)=w_0+(1-a) w_a$ \cite{Chevallier:2000qy,Linder:2002et},  the model predicts 
$w_0\simeq-1.04$ and $w_a\simeq -0.02$  \cite{Maggiore:2013mea}, consistent with the Planck data
\cite{Ade:2013zuv}, and on the phantom side.  This should be compared with models such as that of  ref.~\cite{Deser:2007jk}, which involves an arbitrary function $f(\iBox R)$, which can be chosen so to reproduce any expansion history. The model
(\ref{modela2}) is therefore very predictive, and passes remarkably the  non-trivial test of giving an equation of state consistent with the existing limits.\footnote{Similar interesting cosmological consequences follow from
the model (\ref{SNLnoR3}), and will be explored in \cite{Maggiore:2013mea,Dirian:2014ara}.
Of course, as always in model building, there is always a freedom in the choice of the model itself, and in this sense the study of any single model in this class should be considered as an example of the typical  consequences of  nonlocal terms that can be associated to a mass parameter $m$.}

The purpose of the present paper is to continue the investigation of the model defined by \eq{modela2}, addressing in particular the issue of its classical non-linearities by studying the
static spherically symmetric solutions.
A typical issue of  massive gravity theories is that they become non-linear when $r$ is smaller than the Vainshtein radius,  which is parametrically larger than the \Sch radius $r_S$ of the source. For instance, in the theory defined by adding a Fierz-Pauli mass term to the Einstein-Hilbert action, one finds that the classical  non-linearities become large below the Vainshtein radius 
$r_V=(GM/m^4)^{1/5}$ \cite{Vainshtein:1972sx,Deffayet:2001uk}.
In the dRGT theory \cite{deRham:2010ik,deRham:2010kj} the strong-coupling energy scale is raised and the corresponding critical distance is lowered, to $r_V=(GM/m^2)^{1/3}$ \cite{ArkaniHamed:2002sp}, which is still of order of 100~pc for   $m={\cal O}(H_0)$ and $M=\msun$. In order to recover the successes of GR at solar system and shorter scales, one must then show that a Vainshtein mechanism is at work, i.e. that the inclusion of classical non-linearities restore continuity with GR at $r\ll r_V$. Explicit examples of this type have indeed been found for the dRGT theory
\cite{Koyama:2011xz,Koyama:2011yg}.

In the nonlocal  model (\ref{modela2}) the situation seems however different. Indeed, linearizing the theory over flat space, one finds that the matter-matter gravitational interaction mediated by this theory is given by~\cite{Maggiore:2013mea}  
\be\label{Seff}
S_{\rm eff} =16\pi G\int \frac{d^{4}k}{(2\pi)^{4}}\, \tilde{T}_{\mu\nu}(-k)\Delta^{\mu\nu\rho\sigma}(k)
\tilde{T}_{\rho\sigma}(k)\, , 
\ee
where
\be\label{Delta}
\Delta^{\mu\nu \rho\s}(k)=\frac{1}{2k^2}\, 
\( \eMR\eNS +\eMS\eNR-\eMN\eRS \) 
+ \frac{1}{6}\, \frac{m^2}{k^2(-k^2+m^2)}\eMN\eRS\, .
\ee
The first term  is the usual GR result due to the exchange of a massless graviton, while the extra term vanishes for $m\ra 0$. Therefore this  theory has no vDVZ discontinuity, and no Vainshtein mechanism is needed to restore continuity with GR. 
Of course, the fact that we do not need a Vainshtein mechanism does not necessarily mean that non-linearities will remain small down to the \Sch radius $r_S$, where also the classical non-linearities of GR become large. So, the purpose of this paper is  to study  static spherically symmetric solutions in this theory, and compare with the corresponding solutions of GR. In our problem we have two independent length-scales, the \Sch radius $r_S$ of the source, and the length-scale $m^{-1}$. To have interesting and viable cosmological applications we must have $m={\cal O}(H_0)$ (indeed, the analysis of  \cite{Maggiore:2013mea} shows that the model generates a dynamical dark energy with the observed value of $\ode$ if we choose $m\simeq 0.67 H_0$). Thus, between these two scales there is a huge separation, $r_S\ll m^{-1}$. At scales $r\sim m^{-1}$ we expect that the nonlocal theory (\ref{modela2}) will differ from GR. Indeed, the motivation for such a model is  just to produce a modification of GR in the far infrared, that could account for the observed acceleration of the Universe.  The main motivation of this paper is to see if, in the region $r_S\ll r\ll m^{-1}$, the theory remains linear and close to GR (while, of course, as $r\ra r_S$, even GR becomes non-linear), and to compute explicitly the deviations from GR in the region 
$r\sim m^{-1}$, where they could become relevant for comparison with structure formation.

The paper is organised as follows. In sect.~\ref{sect:basic} we write down the equations of motion in spherical symmetry.  In sect.~\ref{sect:sol} we solve these equations analytically in the regime
$ r\ll m^{-1}$  using a low-$m$ expansion. In sect.~\ref{sect:rmgg1} we solve them in the region $r\gg r_S$, using the linearization over the Minkowski background, and we show that in the region $r_S\ll r\ll m^{-1}$ the solution overlaps with that found in sect.~\ref{sect:sol}.
The results are confirmed through a numerical analysis in Sect.~\ref{sect:num}. Finally, in sect.~\ref{sect:freedom} we give a discussion of  the radiative and non-radiative degrees of freedom of the nonlocal theory, which is useful for a physical understanding of  the  results obtained. Sect.~\ref{sect:concl} contains our conclusions.

\section{Basic equations}\label{sect:basic}

We look for static spherically symmetric solutions of  \eq {modela2}. As in \cite {Maggiore:2013mea},
we  define
\bees
U&=&-\iBox R\label{defU}\, ,\\
\Smn&=&-U\gmn =\gmn\iBox R\, ,\\
B_{\mu\nu}&=&\frac{1}{2}(\n_{\mu}S_{\nu}+\n_{\nu}S_{\mu})\, ,
\ees
Then the original non-local equation (\ref{modela2}) can be formally rewritten as a system of local equations
for the variables $\gmn$, $S_{\mu}$ and $U$,
\be\label{v2loc1}
\Gmn +\frac{m^2}{3}\, \[U\gmn +\frac{1}{2} (\n_{\mu}S_{\nu}+\n_{\nu}S_{\mu})\]=8\pi G\,\Tmn\, ,
\ee
\be
\Box U=-R\, ,
\ee
and
\be\label{panU}
\frac{1}{2}(\d^{\mu}_{\nu}\Box +\n^{\mu}\n_{\nu})S_{\mu}=-\pan U\, , 
\ee
where the latter equation is obtained taking the divergence of \eq{splitSmn}. Some subtleties involved in this localization procedure will be discussed below.
We write the most general static spherically symmetric metric in the form
\be\label{ds2}
ds^2=-e^{2\alpha(r)} dt^2+e^{2\beta(r)}dr^2 +r^2(d\theta^2+\sin^2\theta\,  d\phi^2)\, .
\ee
%\be\label{ds2}
%ds^2=-B(r) dt^2+C(r)dr^2 +
%A(r) r^2(d\theta^2+\sin^2\theta\,  d\phi^2)\, .
%\ee
Observe that the nonlocal equation
(\ref{modela2}) is  generally covariant. 
Therefore, just as in GR, we can use the invariance under diffeomorphisms to set to one a function $e^{2\mu(r)}$ that otherwise, in the most general spherically symmetric solution,  would multiply the term $r^2(d\theta^2+\sin^2\theta\,  d\phi^2)$, and that indeed must be kept in local massive gravity theories.\footnote{In massive gravity models, where there is both a dynamical metric and a reference metric, 
the assumptions of staticity and of spherical symmetry are not sufficient to put both of them in this form, and in one of them remains a term $2D(r) dt dr$ 
\cite{Hinterbichler:2011tt}.
As discussed in
\cite{Koyama:2011xz,Koyama:2011yg}
in dRGT there are two possible branches of solutions: a branch with $D(r)=0$, which is asymptotically flat, exhibits a vDVZ discontinuity at $r\gg r_V$, and recovers GR at $r\ll r_V$, thereby giving an explicit example of the Vainshtein mechanism; and a branch where $D(r)$ is a non-vanishing function, which corresponds to  a  Schwarzschild-de~Sitter solution. In our case, however, there is no reference metric, and for a static and spherically symmetric source 
we can set $D(r)=0$.}
We  use the labels $(0,1,2,3)$ for the   indices $(t,r,\theta,\phi)$ and 
we denote $df/dr$ by $f'$.
The non-vanishing Christoffel symbols in the metric (\ref{ds2}) are
\begin{eqnarray}
\begin{array}{lll}
\Gamma^{0}_{01}=\a'\, ,&\Gamma^{1}_{00}=e^{2(\a-\b)}\a' \, ,
&\Gamma^{1}_{11}=\b'\\
\Gamma^{1}_{22}=-re^{-2\b}\, ,&\Gamma^{1}_{33}= -re^{-2\b}\sin^2\theta \, ,
&\Gamma^{2}_{12}=1/r\\
\Gamma^{2}_{33}=-\sin\theta\cos\theta\, ,&
\Gamma^{3}_{13}=1/r\, ,&\Gamma^{3}_{23}=\cos\theta/\sin\theta\, ,
\end{array}
\end{eqnarray}
%\begin{eqnarray}
%\begin{array}{lll}
% \Gamma^{0}_{01}=B'/(2B)
%&\Gamma^{1}_{00}=B'/(2C)
%&\Gamma^{1}_{11}=C'/(2C)\\
%\Gamma^{1}_{22}=-r/C 
%&\Gamma^{1}_{33}=-(r\sin^2\theta)/C
%&\Gamma^{2}_{12}=1/r\\
%\Gamma^{2}_{33}=-\sin\theta\cos\theta
%&\Gamma^{3}_{13}=1/r
%& \Gamma^{3}_{23}=\cos\theta/\sin\theta
%\end{array}
%\end{eqnarray}
plus those related by the symmetry 
$\Gamma_{\mu\nu}^{\rho}=\Gamma_{\nu\mu}^{\rho}$.
Using these expressions we can compute $B_{\mu\nu}$.
In a spherically symmetric spacetime, for symmetry reasons it is clear that $S_2=S_3=0$, and $S_0=S_0(r)$, $S_1=S_1(r)$. Then, the non-vanishing components
of $B_{\mu\nu}$ are
\begin{eqnarray}
\begin{array}{lll}
B_{00}=-e^{2(\a-\b)}\a' S_1\, ,&B_{11}=S'_1-\b'S_1\, ,&
B_{22}=r e^{-2\b}S_1 \\
B_{33}=r e^{-2\b}\sin^2\theta \, S_1\, ,&
B_{01}=(1/2)S'_0-\a' S_0&B_{10}=B_{01}\, .
\end{array}
\end{eqnarray}
To compute $S_0$ and $S_1$ we use \eq{panU}.
The equations with $\nu=2,3$ are automatically satisfied by $S_2=S_3=0$. Setting  $\nu=0$ gives an equation that only involves $S_0$,
\be
\left\{ \pa_r-\[\b'-\a'-(2/r)\]\right\} \(\pa_r-2\a'\)S_0(r)=0\, ,
\ee
which has the solution $S_0=0$. In principle it also admits other solutions. For instance, the non-vanishing solution of $(\pa_r-2\a')S_0=0$ is $S_0=c_0 e^{2\a(r)}$, with $c_0$ a constant. However, the correct solution is uniquely specified by the condition that $S_0$ must be equal to zero when $U=0$, since for $U=0$ we have $\Smn=0$, which is already trivially transverse, and a non-vanishing $S_{\mu}$ is in this case a spurious solution. Such spurious solutions typically arise when writing the original nonlocal equation as a system of local differential equations, introducing auxiliary fields such as, in our case, $U$ and $S_{\mu}$
\cite{Koshelev:2008ie,Koivisto:2009jn,Barvinsky:2011rk,Deser:2013uya,Foffa:2013vma,Foffa:2013sma} (see also the discussion in sect.~\ref{sect:freedom}). Thus, in our case we only retain the solution $S_0=0$, and the only non-vanishing auxiliary fields are $U(r)$ and $S_1(r)$.
Taking the $\nu=1$ component of \eq{panU} we get
\be\label{eq1}
S_1''+\(\a'-3\b'+\frac{2}{r}\)S_1'-
\(\b''-2\b'^2+\a'^2+\a'\b'+\frac{2}{r}\b'+\frac{2}{r^2}\)S_1=-e^{2\b}U'\, .
\ee
The equation for $U$ is obtained from \eq{defU}, written in the form $\Box U=-R$. 
In general, on a scalar function $U$, $\Box U=(-g)^{-1/2}\pam[\sqrt{-g} g^{\mu\nu}\pan U]$. However, on a function $U(r)$ in a diagonal metric such as  (\ref{ds2}), 
$\Box  U=(-g)^{-1/2}\pa_r[\sqrt{-g} g^{rr}\pa_r U(r)]$ is just a covariant Laplacian.
In the metric (\ref{ds2}) this gives
\be\label{eq2}
e^{-2\b}\[ U''+ \(\a'-\b'+\frac{2}{r}\) U'\]=-R\, ,
\ee
where
\be\label{expliR}
R= \frac{2}{r^2}-e^{-2\b}\[\frac{2}{r^2}+\frac{4}{r}(\a'-\b')+ 2(\a''+{\a'}^2-\a'\b')\]\, .
\ee
Finally, the system of equations for the four functions $\{\a,\b,U,S_1\}$ is completed by taking any two independent components of the nonlocally modified Einstein equations (\ref{modela2}). Writing
$ \(\gmn \iBox_g R\)^{\rm T}\equiv S_{\mu\nu}^T=\Smn-B_{\mu\nu}$, \eq{modela2} can be rewritten as
\be\label{Rmn}
\Rmn =8\pi G\( \Tmn-\frac{1}{2}T\gmn \)
-\frac{m^2}{3}\(B_{\mu\nu}-\frac{1}{2}B\gmn-U\gmn\)\, ,
\ee
where
\be\label{eqB}
B\equiv\gMN B_{\mu\nu}=e^{-2\b}\[ S_1'+\(\a'-\b'+\frac{2}{r}\)S_1\]\, .
\ee
We now study these equation in the region outside the source, where $\Tmn=0$.
Let us recall that in GR one typically takes the combinations 
$e^{2(\b-\a)}R_{00}+R_{11}$ and $R_{22}$ (see e.g. \cite{Carroll:1997ar}). 
In vacuum this gives $e^{2(\b-\a)}R_{00}+R_{11}=0$ and $R_{22}=0$. The former equation gives $\a=-\b$ and then the latter gives a differential equations for $\a$. In our nonlocal theory, using \eq{Rmn} we get instead
\bees
\frac{2}{r}(\a'+\b')&=&\frac{m^2}{3}\[ (\a'+\b')S_1-S'_1\]\, ,\label{eq3}\\
1+e^{-2\b}\[ r(\b'-\a')-1\] &=&
\frac{m^2r^2}{6}\left\{ 2U +e^{-2\b}
\[S_1'+(\a'-\b')S_1\]\right\}\, . \label{eq4}
\ees
Observe, from \eq{eq3}, that now $\a\neq -\b$, unless $S_1$ is constant.
Equations (\ref{eq1}), (\ref{eq2}), (\ref{eq3}) and (\ref{eq4}) provides four differential equations for the four functions $\a(r)$, $\b(r)$, $U(r)$ and $S_1(r)$.

Finally, it is convenient to trade $S_1$  for a field $V(r)$ defined by $S_1(r)=e^{\b}rV(r)$, and work with the four dimensionless functions
$\a(r)$, $\b(r)$, $U(r)$ and $V(r)$. In terms of $V$ the final form of our equations is\bees
&&\hspace*{-1.2cm}
rV''+[4+r(\a'-\b')]V'+
\(\a'-\b'-r{\a'}^2\) V =-e^{\b}U'\, ,\label{eqfull1}\\
&&\hspace*{-1.2cm}
r^2U''+[2r+(\a'-\b')r^2 ]U'=-2e^{2\b}+2\[1+2r(\a'-\b')+ r^2(\a''+{\a'}^2-\a'\b')\],
\label{eqfull2}\\
&&\hspace*{-1.2cm}
\a'+\b'=-\frac{m^2r^2}{6} e^{\b} \[ V'+\(\frac{1}{r}-\a'\)V\]\, ,\label{eqfull3}\\
&&\hspace*{-1.2cm}
1+e^{-2\b}\[r(\b'-\a')-1\]=
\frac{m^2r^2}{6}\left\{
2U+e^{-\b} \[ rV'+\(r\a'+1\)V\]
\right\}\, .\label{eqfull4}
\ees
Observe that replacing $S_1$ by $V$ eliminates the term $\b''$ from \eq{eq1}, and therefore from the whole system of equations.

\section{Solution for  $r\ll m^{-1}$}\label{sect:sol}

We now study \eqst{eqfull1}{eqfull4} in the region $mr\ll 1$, performing a low-$m$ expansion. We assume that in the limit $mr\ra 0$ the terms  on the right-hand side of \eqs{eqfull3}{eqfull4} are negligible, and we will then check a posteriori the self-consistency of the assumption. 
Consider first the equations in the external region, where $\Tmn=0$.
In this case, to lowest order
\eqs{eqfull3}{eqfull4} reduce to their standard GR form, whose solution is given by
\be\label{solAB}
\a(r)=\frac{1}{2}\ln\(1-\frac{r_S}{r}\)\, ,\qquad 
\b(r)=-\a(r)\, .
\ee
Plugging these expressions into \eq{eqfull2} we get
\be\label{UVgen}
U(r)=u_0-u_1\ln\(1-\frac{r_S}{r}\)
\, ,
\ee 
with $u_0, u_1$ some constants, that parametrize the solution of the associated homogeneous equation. Observe that the inhomogeneous solution vanishes since the right-hand side of \eq {eqfull2} is zero on the unperturbed \Sch solution (as it is also obvious from the fact that it is just $r^2R$, and the Ricci scalar $R$ vanishes on the \Sch solution).
The choice of homogeneous solution is a delicate point that requires some discussion.
As discussed in detail in \cite{Maggiore:2013mea,Foffa:2013vma,Foffa:2013sma}, this issue is related to the fact that, in order to complete the definition of the nonlocal model (\ref{modela2}), we must specify what we actually mean by $\iBox$. In general, an equation such as $\Box U=-R$ is solved by
\be\label{Uhom}
U(x)=-\iBox R =U_{\rm hom}(x)-\int d^4x'\, \sqrt{-g(x')}\, G(x;x') R(x')\, ,
\ee
where $U_{\rm hom}(x)$ is any solution of $\Box U_{\rm hom}=0$ and $G(x;x')$ is any a Green's function of the $\Box$ operator. To define our nonlocal integro-differential equation we must specify what definition of $\iBox$ we use, i.e. we must specify the Green's function and the solution of the homogeneous equation. In our static setting the definition of 
$G(x;x') $ is irrelevant, since if $R(x')$ and $\sqrt{-g(x')}$ are independent of $t'$, all possible definitions the above equation reduce to
\be
U(\vx)=U_{\rm hom}(\vx)-\int d^3x'\, \sqrt{-g(\vx')}\, G_L(\vx;\vx') R(\vx')\, ,
\ee
where $G_L(\vx;\vx')=\int dt'  G(x;x')$ is the Green's function of the covariant Laplacian. Still, the definition of the $\iBox$ operator is  completed only once we specify $U_{\rm hom}(\vx)$. In general, this will be fixed by the boundary conditions of the specific problem that we consider, which in our static case
will therefore fix  $u_0$ and $u_1$.
Similarly, we must specify  the homogeneous solution associated to the definition of $S_{\mu}$,
\eq{panU}, which is equivalent to completely define  the nonlocal operation of taking  the transverse part. In other words, $u_0$ and $u_1$ (and the similar constants that characterize the homogeneous solution of $S_{\mu}$)
are not parameters that can be varied and that classify all possible classical solutions of a given theory. Rather, they are fixed once and for all by the definition of the original nonlocal theory and the boundary conditions of the problem at hand. 

In particular, a non-vanishing value of $u_0$ corresponds to introducing a cosmological constant term in the theory.  Indeed, denote by $U_{\rm old}$ and $U_{\rm new}$ two different definitions of $U$ related by  $U_{\rm new}=U_{\rm old}+u_0$. Then
the nonlocal theory using the definition $U_{\rm new}$,  
\be\label{modelnew}
\Gmn +\frac{m^2}{3}\, \(\gmn U_{\rm new}\)^{\rm T}=8\pi G\,\Tmn\, ,
\ee
is equivalent to 
\be
\Gmn +\frac{m^2}{3}\, \(\gmn U_{\rm old}\)^{\rm T}=8\pi G\,\Tmn -\frac{m^2}{3}u_0\gmn\, ,
\ee
and is therefore the same as the old theory, in which we add to the right-hand side a cosmological constant $\Lambda= -(1/3)m^2 u_0$. In this paper we consider the nonlocal model defined by setting $u_0=0$. For such a model, our aim is to show that the static spherically symmetric solutions of the theory reduce to the \Sch solution of GR as $m\ra 0$. Of course, if we switch on $\Lambda$, we should rather show that they approach the corresponding Schwarzschild-dS or (depending on the sign of $\Lambda$) Schwarzschild-AdS solutions. The appropriate choice of $u_1$ is more subtle. We will for the moment keep it generic, and we will later see how it can be uniquely determined. We therefore write
\be\label{USol}
U(r)=-u_1\ln\(1-\frac{r_S}{r}\)
\, ,
\ee 
We now plug these expressions for $\a,\b$ and $U$  into \eq{eqfull1}, and we get
\be\label{solV}
V(r)=\frac{u_1}{6 r^{5/2}(r-r_S)^{1/2}}
\[3 r_S^3-2 r_S^2 r -r_S r^2+2 (r^3-r_S^3)\log\frac{r-r_S}{r_S}
-2 r^3 \log\frac{r}{r_S}\]\, ,
\ee
plus the solutions of the corresponding homogeneous equation, that  we set to zero on the ground that, when  $U=0$, we must have $V=0$, since in this case $U\gmn=0$ and there is no transverse part to extract. Observe that, for $r\gg r_S$, this expression reduces to
$V(r)\simeq -u_1r_S/(2 r)$. We have therefore obtained the solution for $\a,\b$ $U$ and $V$ to zero-th order in $m$.
To get the first correction to $\a$ and $\b$ we plug these expression for $U,V$  back into \eqs{eqfull3}{eqfull4} and solve them. In principle this can be done for generic $r$, as long as $r\ll m^{-1}$, but the resulting solutions, involving polylog functions,  are quite long and not very illuminating, so we we write down the correction term only in the limit $r_S\ll r$. 
Since we are treating $mr$ perturbatively, we are then actually studying the solution in the  region $r_S\, \lsim\, r\ll m^{-1}$. 

The region $r_S\ll r\ll m^{-1}$ is particularly interesting in view of the fact that, in typical massive gravity theories, the classical theory becomes non-linear below a Vainshtein radius $r_V$ parametrically larger than $r_S$. Studying the solution in this region  allows us to investigate whether the same phenomenon happens in the nonlocal theory.
In this regime we can use the zero-th order expressions for $\a$, $\b$, $U$ and $V$ and expand them to leading order in $r_S/r$.
Plugging these  expressions on the right-hand side of \eq{eqfull3}, to leading order in $r_S/r$  we still find\footnote{Keeping also the next-to-leading term we get
$\a'+\b'=-m^2r_S^2u_1/(6r) $. Repeating the analysis performed below, we find that the term 
proportional to $m^2r$ inside the logarithm  in \eq{abfirst} becomes
$m^2r[1+{\cal O}(r^{-1}\log r)]$. This correction is therefore negligible at large $r$.}
\be
\a'+\b'=0\, ,%-\frac{m^2r_S^2u_1}{6r}\, ,
\ee
and, to first non-trivial order, \eq{eqfull4} becomes
\be
2r\b'-1 =-e^{2\b}+e^{2\b} \frac{u_1m^2r_Sr}{3}\, .
\ee
Since we are computing the $m^2$ correction only in the limit $r\gg r_S$, in the term proportional to $m^2$ we can approximate
$e^{2\b}\simeq 1$, so
\be
2r\b'-1\simeq -e^{2\b}+\frac{u_1m^2r_Sr}{3}\, .
\ee
To the order at which we are working,  the solution can be written in the form
\be\label{abfirst}
\a(r)=
-\b(r)=\frac{1}{2}\ln\(1-\frac{r_S}{r}-\frac{u_1m^2 r_S}{6} r\)\, .
\ee
Thus,  in the region $r_S\ll r\ll m^{-1}$,
\be\label{Afinal}
A(r)\equiv e ^{2\a}=1-\frac{r_S}{r}\(1+\frac{u_1m^2 r^2}{6}\)\, ,
\qquad B(r)\equiv e^{2\b(r)}=1/A(r)\, .
\ee
%Finally, we can compute for completeness the first corrections to $U(r)$. Plugging
%\eq{abfirst} into \eq{eqfull2} we get
%\be
%r^2 U''+\[ \frac{r(2r-r_S)}{r-r_S}+\frac{u_1m^2r_Sr^3(2r_S-r)}{6(r-r_S)^2}\] U'=
%-\frac{u_1 m^2 r_S r}{r-r_S}\,.
%\ee
%
The above result shows that our perturbative procedure is self-consistent, since the corrections to linearized theory are indeed small, as long as $mr\ll 1$. It is clear that the procedure can be iterated, obtaining  a systematic expansion  in the small parameter $(mr)^2$.
This should be contrasted with what happens in
massive gravity, when one considers the Einstein-Hilbert action plus a Fierz-Pauli mass term. Then  the analogous computation gives, to first order in the non-linearities, \cite{Vainshtein:1972sx,Hinterbichler:2011tt}
\be\label{AFP}
A(r)=1-\frac{4}{3}\frac{r_S}{r}\(1-\frac{r_S}{12m^4r^5}\)\, .
\ee
The factor $4/3$ in front of $r_S/r$ is due to the extra contribution coming from the exchange of the helicity-0 graviton, and  gives rise to the vDVZ discontiuity. In contrast, no vDVZ discontinuity is present in
\eq{Afinal}. Furthermore the correction terms are crucially different. In 
\eq{AFP} the correction  explodes at low $r$, i.e. for $r$ below the 
Vainshtein radius  $r_V=(GM/m^4)^{1/5}$. In \eq{Afinal}, in contrast, the correction becomes smaller and smaller as $r$ decreases, and perturbation theory is valid at all scales $r\ll m^{-1}$, until we arrive at $r\sim r_S$ where also GR becomes non-linear.
In summary:

\begin{itemize}

\item In the nonlocal theory defined by \eq{modela2} there is no vDVZ discontinuity.
This confirms the result that was found in  \cite{Maggiore:2013mea} by expanding over flat space. 

\item The linearized expansion is fully under control for all distances $r_S\ll r\ll m^{-1}$. Contrary to (local) massive gravity theories, the classical theory stays linear for all distances down to $r\sim r_S$, where eventually also the usual GR non-linearities show up. For $r_S\ll r\ll m^{-1}$ the corrections to GR are actually smaller and smaller as $r$ decreases, the classical theory never becomes strongly coupled, and  recovers  all successes of GR at the solar system and lab scales.

\end{itemize}

The $m^2$ expansion discussed in this section allowed us to obtain perturbatively the solution in the region $r\ll m^{-1}$. As $r$ approaches $m^{-1}$, the corrections become of order one and the small-$m$ expansion breaks down. Furthermore we have written explicitly the correction terms only in the limit $r\gg r_S$. However, this is not due to an intrinsic limitation of the perturbative expansion but is only done for simplicity, since the full expressions are somewhat long. In any case, we found that the corrections are proportional to $m^2r^2$, and becomes smaller and smaller as $r$ decreases toward the horizon, so the terms that we have omitted are in fact negligible even close to the horizon (as we will also check in sect.~\ref {sect:num} comparing the perturbative solutions (\ref {Afinal}) to the result of the numerical integration). Thus the  results of this section are a good approximation to the exact solution in the whole range $r_S\,\lsim\, r\ll m^{-1}$.

\section{Solution for $r\gg r_S$. The Newtonian limit}\label{sect:rmgg1}

The solution in the region $r\gg r_S$, with no limitation of the parameter $mr$, can be obtained with a different expansion, namely
considering the effect of the source  as a perturbation of  Minkowski space, adapting the standard analysis performed in GR to recover the Newtonian limit. This will allow us to obtain analytically the solution in the region
$mr\,\gsim\, 1$, which is not accessible to the low-$m$ expansion. Furthermore, in the region
$r_S\ll r\ll m^{-1}$ both the low-mass and the Newtonian expansions are valid, and  therefore we can match the solutions. We will then confirm the validity of these expansions with the numerical integration in sect.~\ref {sect:num}.

Thus, in this section
we start from a background  $\gbmn=\emn$, $\bar{U}=0$ and $\bar{S}_{\mu}=0$, and  $\Tmn=0$, and we perturb it adding the energy-momentum tensor of a localized source. We limit ourselves to a static non-relativistic source, in which case $\d T_{00}=\rho$, while $\d T_{0i}$ and $\d T_{ij}$  vanish. We are interested in the scalar perturbations so, using the Newtonian gauge, the perturbed metric can be written as
\be\label{ds2Newt}
ds^2 =  -(1+2 \Psi) dt^2 + (1 + 2 \Phi) d\vx^2\, .
\ee
We also expand the auxiliary fields as
$U=\bar{U}+\d U$, $S_{\mu}=\bar{S}_{\mu} +\d S_{\mu}$.
Since the background values $\bar{U}=\bar{S}_{\mu}=0$, we simply write the perturbations as
$U$ and $S_{\mu}$, keeping however in mind  that they are first-order quantities, just as $\Phi$ and $\Psi$. Furthermore, for a static source we necessarily have $S_0=0$, as before, both for the background and the perturbation. The vector $S_i$ can instead be decomposed as usual as  $S_i=S_i^{\rm T}+\pa_i  S$ where $S_i^{\rm T}$ is a transverse vector,  $\pa_iS_i^{\rm T}=0$, which only contributes to vector perturbations, while $S$ is a scalar. 
Since we are studying the scalar sector we  only retain $S$, and we write $S_i=\pa_i S$. Thus, a static source induces  scalar  perturbations  which are described  by the four functions $\Psi,\Phi, U$ and $S$. Observe that we do not need to restrict to spherical symmetry. The vanishing of $S_0$ is just a consequence of the fact that the source is static, so in this problem nothing depend on time, and $\pa_0U=0$, so  
\eq{panU} with $\nu=0$ is a homogeneous equation that has the solution $S_0=0$.

Observe also that the radial coordinate used in this section is different from that used in 
sect.~\ref{sect:sol}, since we are in a different gauge. In fact, if
we linearize  \eq{ds2} we get
\be\label{ds2ablin}
ds^2=-(1+2\alpha) dt^2+(1+2\beta)dr^2 +r^2(d\theta^2+\sin^2\theta\,  d\phi^2)\, .
\ee
This expression is not in the Newtonian gauge. In the Newtonian gauge the factor $(1+2\Psi)$ multiplies the whole term $d\vx^2$, while in \eq{ds2ablin} the factor $(1+2\b)$ only multiplies  $dr^2$.
For clarity, we will continue to denote by $r$ the radial coordinate of the metric (\ref {ds2}), while we denote by $r_N$ the radial coordinate in the Newtonian gauge, so
\eq{ds2Newt} reads
\be\label{ds2rN}
ds^2 =  -(1+2 \Psi) dt^2 + (1 + 2 \Phi) 
\[ dr_N^2 +r_N^2(d\theta^2+\sin^2\theta\,  d\phi^2)\]
\, .
\ee
As discussed above, in the region $r_S\ll r\ll m^{-1}$ both the low-mass expansion of the previous section and the Newtonian expansion of this section hold, and we can therefore match the results. In order to perform the matching, we will however need the relation between the two coordinates, as well as between $\a,\b$ and $\Psi,\Phi$
in this regime. This is easily found observing that, when $r_S\ll r\ll m^{-1}$, the full 
Schwarzschild-like metric (\ref{ds2}) can be linearized and written as in \eq{ds2ablin}. 
We then rewrite the metric (\ref{ds2rN}) as
\be
ds^2 =  -(1+2 \Psi) dt^2 + (1 + 2 \Phi) 
 \(\frac{dr_N}{dr}\)^2 dr^2+(1 + 2 \Phi) r_N^2(d\theta^2+\sin^2\theta\,  d\phi^2)
\, .
\ee
Comparing with \eq{ds2ablin} we see that $r=(1+\Phi)r_N$ and 
$(1+\Phi)dr_N/dr=1+\b$. Inserting here $r_N=(1+\Phi)^{-1}r$  we get $\b=-r\Phi'/(1+\Phi)$ which, to the linearized order at which we are working, is equivalent to $\b=-r\Phi'$ (and, since $r\Phi'$ is already a first-order quantity, we do not need to distinguish $r$ from $r_N$ here). In summary,
in the overlapping region $r_S\ll r\ll m^{-1}$ we can compare the results of the two approaches, using the relations
\be\label{relaz}
r=(1+\Phi)r_N\, ,\qquad \a=\Psi\, ,\qquad \b=-r\Phi'\, .
\ee
After these preliminary remarks, we perform the actual linearization.
We write \eq {modela2} in the form
\bees\label{eqG}
\Gmn +\frac{m^2}{3}\, \(\gmn U\)^{\rm T}&=&8\pi G\,\Tmn\, .\\
\Box U &=&-R\, .\label{boxU}
\ees
Linearizing the $(00)$ components
of \eq{eqG} and setting to zero all time dependences, as appropriate for a static source, we get
\be\label{n2Phi}
\n^2\Phi+\frac{m^2}{6} U=-4\pi G\rho\, .
\ee
Observe that the Laplacian is in principle with respect to the coordinate $r_N$, but since 
all quantities in \eq{n2Phi} are first-order in the perturbations, we can equivalently use $r$. The same will be true for all other equations below.
The $(0i)$ component of \eq{eqG} vanishes identically on time-independent perturbations. The linearization of the $(ij)$ equation gives, setting again to zero all time derivatives,
\be\label{linij}
-\d_{ij} \n^2(\Phi + \Psi)+\pa_i\pa_j(\Phi+\Psi)-\frac{m^2}{3} (U\d_{ij}+\pa_i\pa_jS)=0\, ,
\ee
where, since we are working to linearized order  over Minkowski space, we are free to write all spatial indices as lower indices, and we have used the fact that, for a Newtonian source, $T_{0i}$ can be neglected. Applying to this equation the projector
$ ( \n^{-2}\partial_i \partial_j  - \frac{1}{3} \delta_{ij} )$ to obtain the traceless part,
we get
\be\label{PhipiuPsi}
\n^2\(\Phi+\Psi-\frac{m^2}{3} S\)=0\, .
\ee
One might be tempted to rewrite this equations as $\Phi+\Psi-(m^2/3) S=0$, but this would not be correct. In general, if a function $f$ satisfies $\n^2 f=0$ over {\em all}  of space, and we further impose the boundary condition that $f$ vanishes at infinity, then $f=0$. However, the equations that we are writing in this section are only valid for $r\gg r_S$. Of course, from the fact that a function $f$ satisfies $\n^2f=0$ at large $r$ we cannot conclude that $f$ itself is identically zero at large $r$. Indeed, any function that, 
at large $r$, approaches the form $f(r)=c_0+c_1r_S/r$ satisfies $\n^2f=0$ at large $r$. In our problem, the constant term $c_0$ is eliminated requiring that the functions $\Phi$, $\Psi$ and $S$ vanish at infinity. We remain however with the possibility of a $1/r$ term. Thus,
\eq{PhipiuPsi} only implies that, at $r\gg r_S$,
\be
\Phi+\Psi=\frac{m^2}{3} S +c_1\frac{r_S}{r}\, ,
\ee
for some constant $c_1$, which can be determined by matching the solution with those found in sect.~\ref {sect:sol}  for $r\ll m^{-1}$, as we will do below.
Taking the trace of 
\eq {linij} and combining it with \eqs{PhipiuPsi}{n2Phi} we get 
\be\label{Un2S}
U=-\n^2S\, .
\ee
This completes the linearization of the nonlocally modified Einstein equation. To complete our system of equations we must also linearize \eqs{panU}{boxU}.  The linearization of \eq{boxU} gives
\be\label{UPhiPsi}
\n^2U=\n^2(2\Psi+4\Phi)\, .
\ee
Again, this equation is only valid at $r\gg r_S$ and only implies that, in such a region,
\be\label{UPhiPsic2}
U=2\Psi+4\Phi +c_2\frac{r_S}{r}\, ,
\ee
for some constant $c_2$.
The linearization of \eq{panU} with $\nu=0$ is identically zero, while that with $\nu=i$ gives a combination of the previous equations. In conclusion, 
\eqsss{n2Phi}{PhipiuPsi}{Un2S}{UPhiPsi} are four equations for the four functions
$\Phi,\Psi$, $U$ and $S$.  Using \eqss{PhipiuPsi}{n2Phi}{Un2S} to transform the right-hand side of
\eq{UPhiPsi} we get
\be
(\n^2+m^2) U=-8\pi G\rho\, .
\ee
This is an inhomogeneous Helmholtz equation, and we can solve it writing
\be\label{Phirhobar}
U (\vx)=-8\pi G\int_V d^3x'\, G(\vx-\vx')\rho(\vx')\, ,
\ee
where 
\be\label{eqnmu}
(\n^2+m^2)G(\vx)=\d^{(3)}(\vx)\, .
\ee
The Green's function of the  inhomogeneous Helmholtz equation is  well known. Writing
$G(r)=-[1/(4\pi r)] f(r)$
one finds
\be
(\n^2+m^2)G(\vx)=\d^{(3)}(\vx) f(0) -\frac{1}{4\pi r} (f''+m^2 f)\, ,
\ee
and therefore $f(0)=1$ and $f''+m^2 f=0$. The most general solution is then
\be
f(r)=\cos (m r) + \b \sin (m r)\, ,
\ee
with $\b$ arbitrary. The corresponding solution for $U$ is 
\be\label{Urhom}
U(\vx)=2G\int d^3x'\, \frac{\rho(\vx')}{|\vx-\vx'|}
\[\cos (m |\vx-\vx'|) + \b \sin (m |\vx-\vx'|)\]\, .
\ee
In  the $r\gg r_S$ limit this becomes
\be\label {U4G}
U(\vx)\simeq \frac{2G}{r}\int d^3x'\, \rho(\vx')
\[\cos (m |\vx-\vx'|) + \b \sin (m |\vx-\vx'|)\]\, .
\ee
In particular, for $\rho(\vx)=M \d^{(3)}(\vx)$, \eq{U4G} gives
\be\label{U4Glarge}
U(r)=\frac{r_S}{r}
\[\cos (mr) + \b \sin (mr)\]\, .
\ee
More generally, even if $\rho(\vx)$ is not a Dirac delta, at distances $r$ much larger than the source size we can write $|\vx-\vx'|\simeq r-\vx'{\bf\cdot}\hatn$, where $\hatn=\vx/r$, so
$\cos (m|\vx-\vx'|)\simeq \cos(mr- m\vx'{\bf\cdot}\hatn)$. For $m={\cal O}(H_0)$, all over the source $m|\vx'|$ is negligibly small with respect to one (and not just with respect to $mr$) and we can  replace
$\cos (m|\vx-\vx'|)$ by $\cos (mr)$. Therefore, at large distances the coefficient of the $1/r$ term  for a generic $\rho(\vx)$ is the same as for  $\rho(\vx)=M \d^{(3)}(\vx)$, just as in GR.

The appropriate Green's function, and therefore the value of $\b$, is fixed by the boundary conditions. In most problems in which the inhomogeneous Helmholtz equation appears, the Green's function is fixed imposing a no-incoming wave boundary condition at infinity, which selects $G(r)=- e^{im r}/(4\pi r)$, i.e. $\b=i$. However, such a boundary condition is not appropriate to our problem, since $U(r)$ is real. In our case, for an extended source, $\b$ must rather be fixed by matching this large distance solution to the solution in the inner source region, as we will discuss below. 

We can now plug this solution for $U$ into \eq{n2Phi}. Using for simplicity
$\rho(\vx)=M \d^{(3)}(\vx)$, we get
\be\label{n2Phicos}
\n^2\Phi=-2\pi r_S \d^{(3)}(\vx)-\frac{ m^2 r_S}{6r}\[\cos (mr) + \b \sin (mr)\]
\, .
\ee
Once again, since the equation only holds at large $r$, we have the freedom of adding
a term proportional to $1/r$ to the solution.  Observing that 
$\n^2[\cos (mr)/r]=-m^2\cos (mr)/r$ and $\n^2[\sin (mr)/r]=-m^2\sin (mr)/r$,
the solution can be written as
\be\label{PhicPhi}
\Phi=\frac{r_S}{2r}\left\{c_{\Phi} +\frac{1}{3}\[\cos (mr)+\b\sin(m r)\]\right\}
\ee
for some constant $c_{\Phi}$. Similarly, plugging \eqs{U4Glarge}{PhicPhi} into
\eq{UPhiPsic2} we get
\be\label{PsicPsi}
\Psi=\frac{r_S}{2r}\left\{c_{\Psi} +\frac{1}{3}\[\cos (mr)+\b\sin(m r)\]\right\}
\ee
where $c_{\Phi}=-(c_2+2 c_{\Psi})$ is a second independent constant. 
To compare with the functions $A(r)$ and $B(r)$ of the previous section 
we use the fact that, in the region $r_S\ll r$, $A(r)=1+2\a(r)$ and
$B(r)=1+2\b(r)$. Using \eq{relaz} we therefore have $A(r)=1+2\Psi$ and $B(r)=1-2r\Phi'$, so
\bees
\hspace*{-5mm}A(r)&=&1+\frac{r_S}{r}\left\{c_{\Psi} +\frac{1}{3}\[\cos (mr)+\b\sin(m r)\]\right\}\, ,\label{AA}\\
\hspace*{-5mm}B(r)&=&1+\frac{r_S}{r}\left\{ c_{\Phi}+\frac{1}{3}\[\cos (mr)+\b\sin(m r)\]
+\frac{mr}{3}\[\sin (mr)-\b\cos(m r)\]\right\}\, .\label{BB}
\ees
 In the limit $mr\ll 1$ from \eqs {PhicPhi}{PsicPsi} we get
\bees
A(r)&=&1+\frac{r_S}{r}
\(c_{\Psi} +\frac{1}{3}+\b mr-\frac{m^2r^2}{6}\)\, ,\\
B(r)&=&1+\frac{r_S}{r}
\(c_{\Phi} +\frac{1}{3}+\frac{m^2r^2}{6}\)\, .
\ees
Matching these expression with the solution (\ref{Afinal}), which is valid for $r\ll m^{-1}$, we get $\b=0$, $c_{\Psi}=-4/3$ and $c_{\Phi}=2/3$. On the other side, comparing the terms $m^2 r^2$, allows us to fix $u_1$ in the solution  (\ref{Afinal}), and  we get $u_1=1$. The latter   result could have also been derived more directly matching the small $mr$ limit of \eq{U4Glarge} to the large $r/r_S$ limit of \eq{USol}.

In conclusion, plugging the value of these constant into the full solutions
(\ref{AA}) and (\ref{BB}) we find that, in the Newtonian limit,
\be
ds^2=-A(r)dt^2 +B(r)dr^2 +r^2(d\theta^2+\sin^2\theta\,  d\phi^2)\, ,
\ee
where
\bees
A(r)&=&1-\frac{r_S}{r}\[1+\frac{1}{3}(1-\cos mr)\]\, ,\label{NewtA}\\
B(r)&=&1+\frac{r_S}{r}\[1-\frac{1}{3}(1-\cos mr)+\frac{1}{3}mr \sin m r\]\, ,\label{NewtB}
\ees
while the auxiliary field $U=-\iBox R$ is given by
\be
U(r)=\frac{r_S}{r}\cos mr \, .\label{NewtU}
\ee

\section{Numerical integration}\label{sect:num}

We now study the equations numerically, in order to confirm the above analytic results. In the numerical analysis it is convenient to trade $U$ for a field $W$ defined by $U=W+2\a$. Then \eqst{eqfull1}{eqfull4} become
\bees
&&\hspace*{-1cm}
rV''+[4+r(\a'-\b')]V'+
\(\a'-\b'-r{\a'}^2\) V =-e^{\b}(W'+2\a')\, ,\label{Weqfull1}\\
&&\hspace*{-1cm}
r^2W''+[2r+(\a'-\b')r^2] W'=2(1-e^{2\b})-4r\b'\, ,
\label{Weqfull2}\\
&&\hspace*{-1cm}
\a'+\b'=-\frac{m^2r^2}{6} e^{\b} \[ V'+\(\frac{1}{r}-\a'\)V\]\label{Weqfull3}\\
&&\hspace*{-1cm}
1+e^{-2\b}\[r(\b'-\a')-1\]=\frac{m^2r^2}{6}\left\{
2W+4\a+e^{-\b} \[ rV'+\(r\a'+1\)V\]
\right\}\, .\label{Weqfull4}
\ees
The advantage of this transformation is that now $\a''$ disappeared from \eq{Weqfull2}. To integrate the equations we need to assign the initial conditions. To this purpose, we take advantage of the fact that we know the zero-th order solution is quite close to the exact solution in the region
$r_S\ll r\ll m^{-1}$. We therefore  choose
a value $r_{\rm in}$ in this region, and  we assign $\a(r_{\rm in})$, $\b(r_{\rm in})$, $U(r_{\rm in})$, $U'(r_{\rm in})$
$V(r_{\rm in})$ and  $V'(r_{\rm in})$ using the zero-th solution
given by \eqss{solAB} {USol}{solV}, setting  $u_1=1$. 
We show for definiteness the results obtained choosing
$r_{\rm in}=200 r_S$ and $m^{-1}=10^3 r_S$, so indeed 
$r_S\ll r\ll m^{-1}$. 

\begin{figure}[t]
\begin{center}
\begin{minipage}{1.\linewidth}
\centering
\includegraphics[width=0.45\columnwidth]{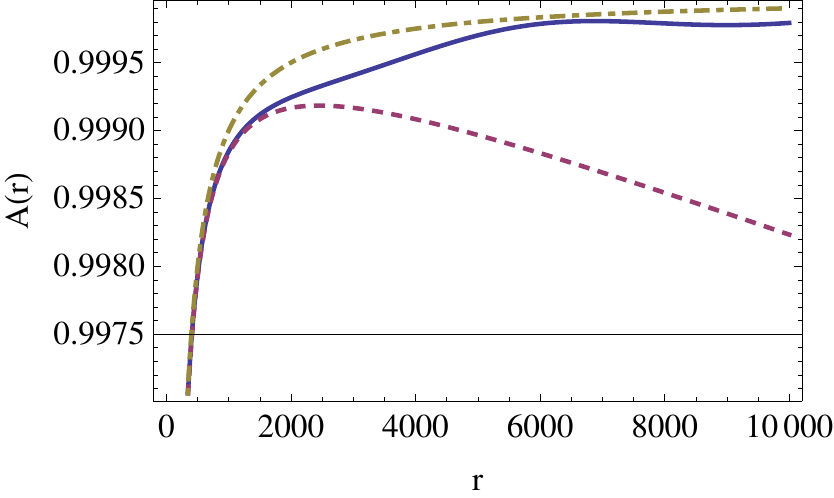}
\includegraphics[width=0.45\columnwidth]{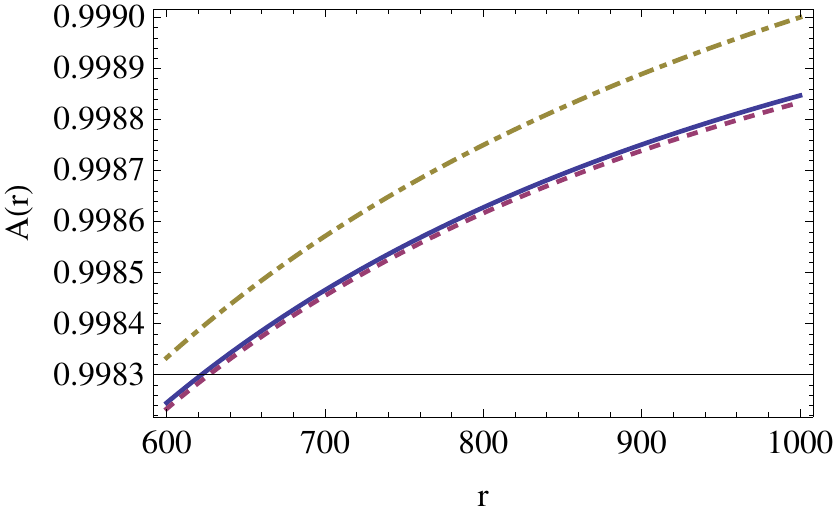}
\end{minipage}
\caption{Left: the function $A(r)$ from the numerical integration (blue solid line), compared with the zero-th order \Sch solution
$A(r)=1-r_S/r$ (brown, dot-dashed)  and with the result of the first-order perturbative low-$m$ expansion
(\ref{Afinal}) (red, dashed). The variable $r$ is measured in units of $r_S$, we set $m^{-1}=10^3$ and we start the integration at $r_{\rm in}=200$.
Right: a zoom-in of the intermediate region  $600<r<1000$, i.e.
$0.6 <mr <1$.  
\label{figA1A2}
}
\end{center}
\end{figure}

\begin{figure}[h]
\begin{center}
\begin{minipage}{1.\linewidth}
\centering
\includegraphics[width=0.45\columnwidth]{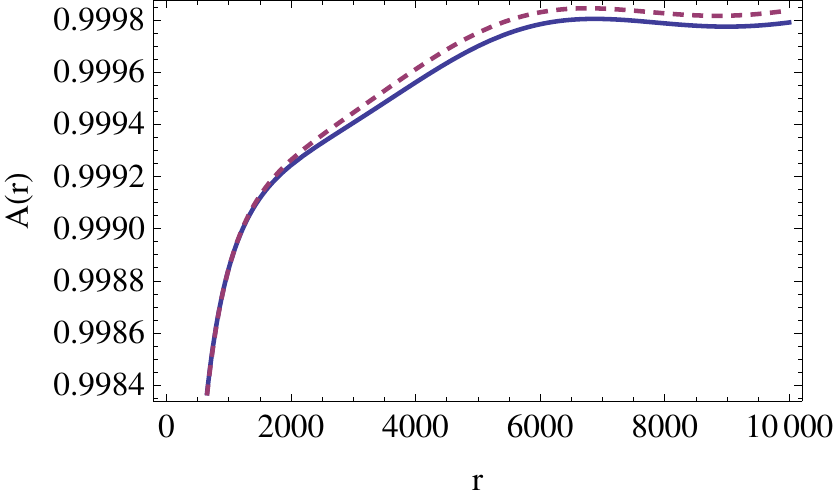}
\end{minipage}
\caption{The function $A(r)$ in the regime $r\gg r_S$  from the numerical integration (blue solid line), compared to the Newtonian solution (\ref{NewtA}) (red, dashed).
\label{figA3}
}
\end{center}
\end{figure}

\begin{figure}[t]
\begin{center}
\begin{minipage}{1.\linewidth}
\centering
\includegraphics[width=0.45\columnwidth]{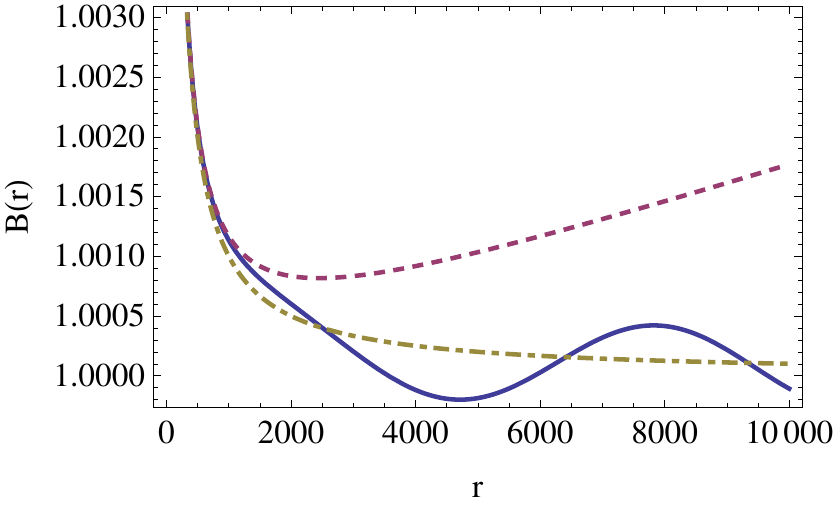}
\includegraphics[width=0.45\columnwidth]{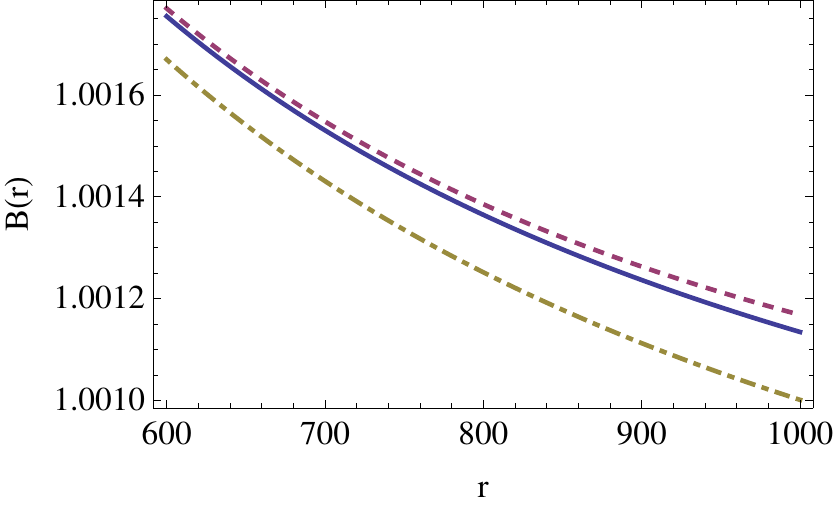}
\end{minipage}
\caption{The same as in fig.~\ref{figA1A2} for the function $B(r)$.
\label{figB1B2}
}
\end{center}
\end{figure}

\begin{figure}[h]
\begin{center}
\begin{minipage}{1.\linewidth}
\centering
\includegraphics[width=0.45\columnwidth]{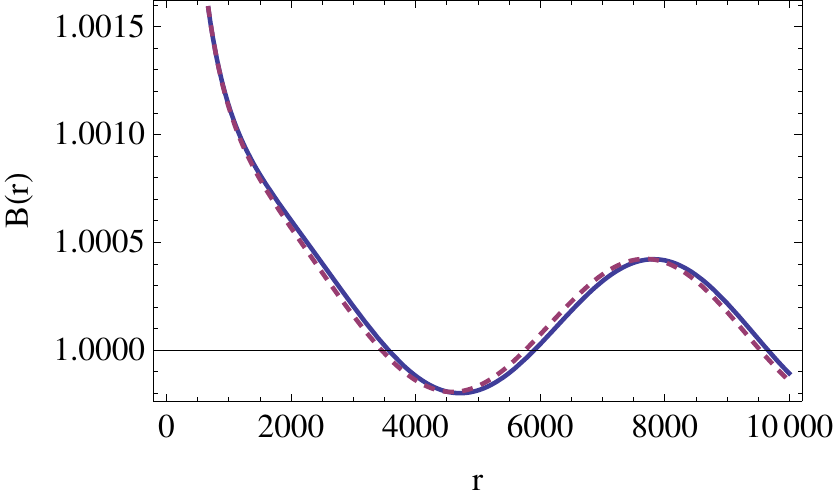}
\end{minipage}
\caption{The same as in fig.~\ref{figA3} for the function $B(r)$.
\label{figB3}
}
\end{center}
\end{figure}

The left panel in Fig.~\ref{figA1A2}  shows the numerical result for the function $A(r)$ (blue solid line)  and compares it with the zero-th order \Sch solution
$A(r)=1-r_S/r$ (brown, dot-dashed) and with the first-order perturbative solution obtained from the low-$m$ expansion,
\eq{Afinal} (red, dashed), over a broad range of values of $r$,
$r_S\leq r <10 m^{-1}$, i.e. $1<r<10^4$ in units $r_S=1$.
The numerical integration confirms that, at $r\ll m^{-1}$, the analytic solutions obtained in a low-$m$ expansion work well, and the first-order correction improves on the zero-th order \Sch solution. 
As $mr$ becomes of order one, the $m=0$ \Sch solution remains relatively close to the numerical result, while  the truncation 
(\ref{Afinal}) goes astray. This is not surprising, since the $m^2r^2$ correction in
\eq{Afinal} is only valid in the regime where it is very small compared to one.
On the right panel  of Fig.~\ref{figA1A2}  we show in more detail the intermediate region  
$0.6 <mr <1$. We see that here the first-order perturbative result improves on the 
zero-th order solution, confirming  the validity of the perturbative expansion. 
In   Fig.~\ref{figA3} we show the function $A(r)$ in the regime $r\gg r_S$  from the numerical integration (blue solid line), compared to the Newtonian solution (\ref{NewtA}) (red, dashed). Again, we see that the analytic solution works well.
Similar results hold for $B(r)$, and are shown  in Figs.~\ref{figB1B2} and \ref{figB3}.

\begin{figure}[th]
\begin{center}
\begin{minipage}{1.\linewidth}
\centering
\includegraphics[width=0.45\columnwidth]{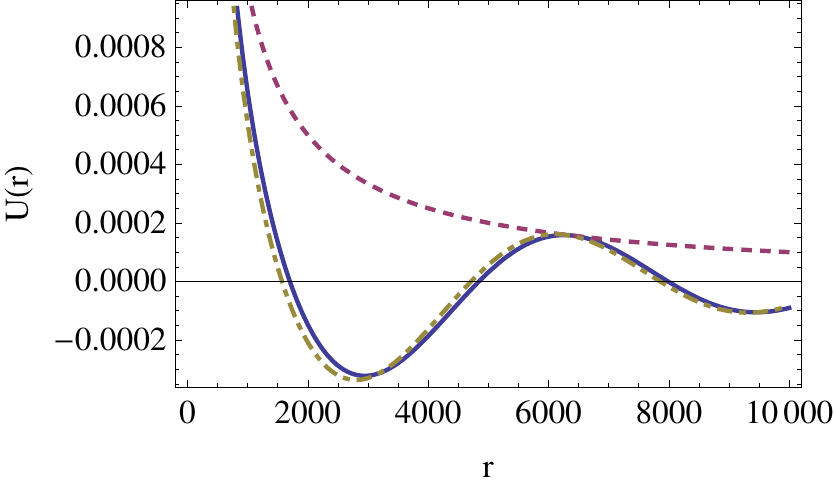}
\includegraphics[width=0.4\columnwidth]{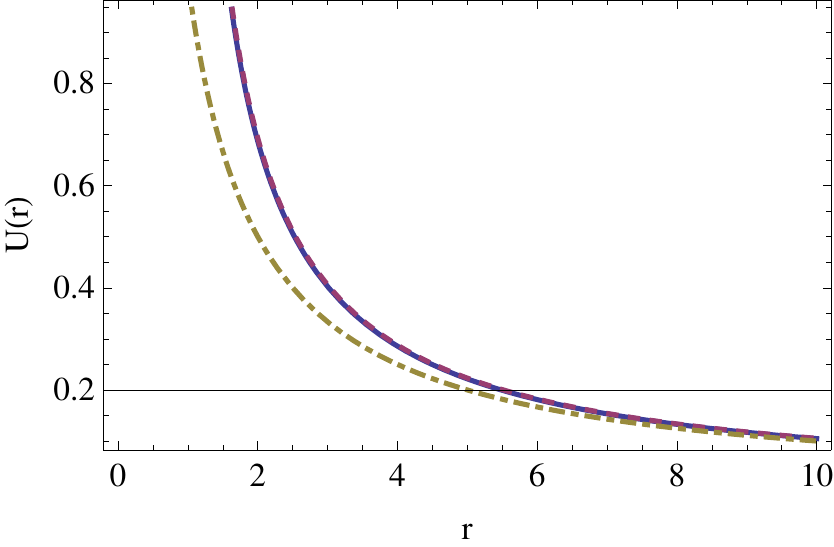}
\end{minipage}
\caption{The numerical solutions for $U$ compared to \eq{USol} with $ u_1=1$ (red dashed line) and to the Newtonian solution (\ref{NewtU}) (brown dot-dashed)
in two different regions, at $rm$ large (right panel) and near the horizon (left panel).
\label{figU1U2}
}
\end{center}
\end{figure}

The numerical solution for $U$ is shown in  Fig.~\ref{figU1U2}. On the left panel we show the numerical integration (blue solid line)  compared to the 
 zero-th order low-$m$ solution (\ref{USol}) (red dashed line) and to the Newtonian solution (\ref{NewtU}) on a large scale that emphasizes the region $mr\gg 1$. Here the Newtonian solutions works well, as expected while, of course, the low-$m$ expansion is not accurate. As we move toward lower values of $mr$ the two curves approach each other. As shown in the right panel, close to the horizon the low-$m$ expansion works extremely well (on the scale of the figure it is indistinguishable from the numerical result) while the Newtonian result becomes less accurate.
These plots confirm that the theory never becomes non-linear in the region 
$r_S\ll r\ll m^{-1}$. The exact numerical solution follows the analytic solution obtained in an expansion in powers of $mr$, until $mr$ becomes of order one. There is no Vainshtein radius $r_V\gg r_S$ below which the low-mass expansion  fails. 

It is also interesting to study the stability under perturbations of the solution that we have found. \Eq{Urhom} and the discussion below \eq{U4Glarge} show  that, if we perturb the source replacing
$\rho\ra \rho +\d\rho$, while still preserving the fact that the source has compact support, no instability develops, and the only effect of $\d \rho$ 
is to  replace the source mass $M$ by the corresponding value $M+\d M$, just as in GR.
Concerning the near-horizon region of a BH solution,  our analytic and numerical results indicate that the corrections to the \Sch solution near the horizon are ${\cal O}(m^2r_S^2)$, which for $m\sim H_0$ is negligibly small, so we do not expect any instability to develop in the  BH quasi-normal modes.

\section{Degrees of freedom of the nonlocal theory}\label{sect:freedom}

To better understand  the meaning of the results obtained, it is useful to discuss what are the radiative and non-radiative degrees of freedom of the theory. 
This issue has already been examined  in   detail  in refs.~\cite{Maggiore:2013mea,Foffa:2013sma} (see also \cite{Maggiore:2014sia} for a related analysis in a similar nonlocal model). However, we find useful to summarize here the main results discussed in the above papers, and compare them with what we have learned above.

Counting the propagating degrees of freedom in a nonlocal theory (or even in a local theory when we transform to nonlocal variables)
involves  some subtleties, of which one must be aware in order  to get  correct results. The simplest example of what can go wrong with nonlocal transformations  is provided by a theory in which, among other fields, also appears a scalar field $\phi$ that satisfies a Poisson equation  $\n^2\phi=\rho$ \cite{Foffa:2013sma}. This field is clearly non-radiative. If, in the classical equation, we set the source $\rho=0$, we simply have $\phi=0$. There are no associated plane waves  freely propagating in empty space associated to this field, and at the quantum level there are no creation and annihilation operators associated to it. However,
if we  define a new field $\tilde{\phi}$ from  $\tilde{\phi}=\iBox\phi$,  the original  Poisson  equation   can be rewritten as  
\be\label{eqtildephi}
\Box\tilde{\phi}=\n^{-2}\rho\equiv\tilde{\rho}\, , 
\ee
so now $\tilde{\phi}$ looks like a propagating degree of freedom. However, for $\rho=0$ our original equation $\n^2\phi=\rho$ only has the solution $\phi=0$. If we want to rewrite it in terms of $\tilde{\phi}$ without introducing spurious degrees of freedom we must therefore supplement \eq{eqtildephi} with the condition that, when $\rho=0$, $\tilde{\phi}=0$. In other words,  the solutions of the associated homogeneous equation $\Box\tilde{\phi}=0$ must be discarded (or, more generally, is uniquely fixed by the boundary conditions of the problem).  Correspondingly, the coefficients
$a_{\vk}, a_{\vk}^{*}$ of the general plane-wave solution of the equations $\Box\tilde{\phi}=0$
cannot be considered as free parameters that,
upon quantization, give rise to the creation and annihilation operators of the quantum theory, and there are no propagating quanta associated to  $\tilde{\phi}$.

A similar situation also appears  in general relativity. Consider  GR linearized over flat space, $\gmn=\emn+\hmn$, and decompose as usual $\hmn$ as
\bees
h_{00}&=&2\psi\, ,\qquad
h_{0i}=\beta_i+\pa_i\gamma\label{h0ibetagamma}\\
h_{ij}&=&-2\phi\d_{ij}+\(\pa_i\pa_j-\frac{1}{3}\d_{ij}\n^2\)\lambda  +\frac{1}{2}(\pa_i\eps_j+\pa_j\eps_i)
 +h_{ij}^{\rm TT}\, ,\label{hijphilambdaeps}
\ees
where $\eps^i$ and $\beta^i$ are transverse vectors, $\pa_i\beta^i=
\pa_i\eps^i=0$, and $h_{ij}^{\rm TT}$ is transverse and traceless, 
$\pa^jh_{ij}^{\rm TT}=0$ and
$\d^{ij}h_{ij}^{\rm TT}=0$. With these variables we can form the gauge-invariant combinations
$\Phi=-\phi-(1/6)\n^2\lambda$ and
$\Psi = \psi -\dot{\gamma}+(1/2)\ddot{\lambda}$, which describe two degrees of freedom in the scalar sector, and the gauge-invariant transverse vector
$\Xi_i=\beta_i-(1/2)\dot{\eps}_i$, which describes two degrees of freedom in the vector sector. These gauge-invariant quantities  are the usual Bardeen's variables specialized to flat space. The remaining degrees of freedom (two in the scalar sector and two in the vector sector) can be set to zero with a gauge transformation, while in the helicity-2 sector
$h_{ij}^{\rm TT}$ is gauge invariant. We have therefore split the 10 components of $\hmn$ into four pure gauge modes and six physical (i.e. gauge-invariant) degrees of freedom. Of course, not all these six degrees of freedom are radiative. To see this, we decompose similarly the energy-momentum tensor as
\bees
T_{00}&=&\rho\, ,\qquad
T_{0i}=\Sigma_i+\pa_i\Sigma\, ,\label{T0i}\\
T_{ij}&=&P\d_{ij}+\(\pa_i\pa_j-\frac{1}{3}\d_{ij}\n^2\)\sigma   +\frac{1}{2}(\pa_i\sigma_j+\pa_j\sigma_i)+\sigma_{ij} \, ,\label{Tij}
\ees
where 
$\pa_i\sigma^i=0$, $ \pa_i \Sigma^i=0$, 
$\pa^i\sigma_{ij}=0$ and $ \d^{ij}\sigma_{ij}=0$. The linearized equations of motion can then be written as \cite{Jaccard:2012ut}
\bees
\n^2\Phi &=&-4\pi G\rho\, ,\label{n2PhiJMM}\\
\n^2\Psi &=&-4\pi G (\rho-2\n^2\sigma)\, ,\label{n2PsiJMM}\\
\n^2\Xi_i&=&-16\pi G \Sigma_i\, ,\\
\Box h_{ij}^{\rm TT}&=&-16\pi G \sigma_{ij}\label{hijsij}\, .
\ees
This shows that   only the tensor perturbations obey a wave equation and are therefore radiative. The gauge-invariant scalar and vector perturbations obey a Poisson equation, and therefore represent physical but non-radiative degrees of freedom.

Now, compare this result with that obtained decomposing $\hmn$ as
\be\label{decomphmn}
\hmn =\hmn^{\rm TT}+\frac{1}{2}(\pam \eps_{\nu}+\pan \eps_{\mu}) +\frac{1}{3}\emn s\, ,
\ee
where $\hmn^{\rm TT}$ is transverse and traceless with respect to the Lorentz indices, $\paM \hmn^{\rm TT}=0$, 
$\eMN \hmn^{\rm TT}=0$, and therefore has five independent components. 
The 10 components of the metric perturbations are therefore split into the five components of $\hmn^{\rm TT}$, the four components of $\eps^{\mu}$, plus the scalar $s$.
Under a linearized diffeomorphism
$\hmn\ra\hmn -(\pam\xin+\pan\xim)$ we have $\eps_{\mu}\ra\eps_{\mu}-\xi_{\mu}$ while the tensor $\hmn^{\rm TT}$ and the scalar $s$ are gauge invariant. We now plug this decomposition into the quadratic Einstein-Hilbert action, 
\be\label{Squadr}
S_{\rm EH}^{(2)}=\frac{1}{2}\int d^{4}x \,
\hmn{\cal E}^{\mu\nu,\rho\sigma}\hrs\, ,
\ee
where  ${\cal E}^{\mu\nu,\rho\sigma}$  is the  Lichnerowicz operator (and we rescaled $\hmn\ra\kappa\hmn$, where $\kappa=(32\pi G)^{1/2}$, in order to have a canonically normalized kinetic term). The result is
\be\label{SEH2nl}
S_{\rm EH}^{(2)}=\frac{1}{2}\int d^{4}x \,\[\hmn^{\rm TT}\Box (h^{\mu\nu })^{\rm TT}
-\frac{2}{3}\, s\Box s\]\, .
\ee
Performing the same decomposition in the energy-momentum tensor, the interaction term 
can be written as
\be\label{Sint2}
S_{\rm int}=\frac{\kappa}{2}\int d^{4}x\, \hmn\TMN 
=\frac{\kappa}{2}\int d^{4}x\,\[ \hmn^{\rm TT}(\TMN)^{\rm TT}+\frac{1}{3}sT\]
\, ,
\ee 
so the equations of motion derived from $S_{\rm EH}^{(2)}+S_{\rm int}$ are
\bees
\Box\hmn^{\rm TT}&=&-\frac{\kappa}{2}\Tmn^{\rm TT}\, ,\label{BoxhTT}\\
\Box s&=&\frac{\kappa}{4}T\, .\label{Boxs}
\ees
At first sight this result is surprising, because it seems to suggest that the five components of the transverse-traceless tensor $\hmn^{\rm TT}$ and the scalar $s$ are all radiative fields. Note that  these degrees of freedom are gauge invariant, so they cannot be  gauged away.\footnote{Observe also that we have {\em not} used linearized gauge invariance to set $\eps^{\mu}=0$ in the action (in which case, one should have been worried that we might have lost the equations obtained performing the variation with respect to $\eps^{\mu}$). Rather, 
$\eps^{\mu}$ disappears automatically from the action, so there is no equation of motion associated to it. The fact that $\eps^{\mu}$ disappears automatically from the action is just a property of  ${\cal E}^{\mu\nu,\rho\sigma} $ when applied to a tensor of the form $\parho\eps_{\sigma}$. Of course, this  property of the ${\cal E}^{\mu\nu,\rho\sigma} $ is just what guarantees the invariance of the quadratic action under linearized  gauge transformations
$\hmn\ra\hmn -(\pam\xin+\pan\xim)$.}
Furthermore, according to \eq{SEH2nl}
the scalar $s$ should be a ghost! 
Of course these conclusions are wrong, and the correct conclusion is the one drawn from 
\eqst{n2PhiJMM}{hijsij}, namely that in GR there are two radiative and four non-radiative degrees of freedom. What went wrong is the following. Eqs.~(\ref{n2PhiJMM})-(\ref{hijsij}) can be inverted, to give $\Phi$, $\Psi$, etc. in terms of the original field $\hmn$. The inversion involves the inverse Laplacian, and is therefore nonlocal in space, but is local in time. In particular this means that there is a one-to-one correspondence between the initial condition assigned on $\hmn$ on a given time slice, and those assigned on $\Phi, \Psi$, etc. on the same time slice. In contrast, when we invert \eq{decomphmn}, we find that the inversion involves the inverse d'Alembertian, and is therefore non-local even in time.  
Thus, this one-to-one correspondence on the initial conditions is lost, and the counting of degrees of freedom goes wrong. In particular, the inversion of \eq{decomphmn} gives
\be\label{defshmn}
s=\(\eMN-\frac{1}{\Box}\paM\paN\)\hmn\, ,
\ee 
and the relation between $s$ and the Bardeen variables $\Phi$ and $\Psi$ is
\be\label{sPhiPsi}
s=6\Phi- 2\iBox\n^2(\Phi+\Psi)\, .
\ee
We see that the situation is exactly the same as that illustrated by \eq{eqtildephi}: the non-radiative field
$\Phi+\Psi$ is transformed into an apparently radiative field $s$ by the nonlocal relation
(\ref{sPhiPsi}), which involves $\iBox$. The bottom-line is that, if in GR we wish to use the decomposition (\ref{decomphmn}) and the fields $s$ and $\hmn^{\rm TT}$ we can do it, provided that we supplement
\eq{Boxs} with the condition that, when $\rho=\sigma=0$, we must have $s=0$, i.e. we must discard the solution of the homogeneous equation $\Box s=0$, since when $\rho=\sigma=0$ we see from
\eqs{n2PhiJMM}{n2PsiJMM} that $\Phi=\Psi=0$
 (and similarly for the component $\hmn^{\rm TT}$ with  helicities $0$
and $\pm 1$). So, again, there are no creation and annihilation operators associated to these fields in the quantum theory, and they cannot appear in external lines nor in loops.\footnote{Observe that, if $s$ could appear on external lines, it would induce vacuum decay processes such as the decay of vacuum into gravitons and would-be ghosts fields $s$. Such diagrams could not be canceled by diagrams where the $s$ lines are replaced by the helicity-0 component of   $\hmn^{\rm TT}$, since they correspond to different final states. See also the more extended discussion in sect.~3.1 of ref.~\cite{Foffa:2013sma}.}

Having understood these simple but important points in the familiar context of GR, we are now well-armed for understanding the situation in the nonlocal theory. As we discussed in sect.~\ref{sect:basic},
using the auxiliary field $U=-\iBox R$ as well as the auxiliary four-vector field $S_{\mu}$ that enters in the extraction of the transverse part in \eq{splitSmn}, \eq{modela2} can be rewritten as
\be\label{loc1}
\Gmn +\frac{m^2}{3}\, \[U\gmn +\frac{1}{2} (\n_{\mu}S_{\nu}+\n_{\nu}S_{\mu})\]=8\pi G\,\Tmn\, ,
\ee
together with
$U=-\iBox R$ and 
\be\label{loc2}
\n^{\mu}S_{\mu\nu}=\frac{1}{2}(\d^{\mu}_{\nu}\Box +\n^{\mu}\n_{\nu})S_{\mu}\, , 
\ee
which is obtained taking the divergence of \eq{splitSmn}. Naively, one would say that 
$U=-\iBox R$ is equivalent to $\Box U=-R$, and therefore the original nonlocal model can be written as a set of local equations for $\gmn, U$ and $S_{\mu}$, given by \eqs{loc1}{loc2} together with $\Box U=-R$. However, it is just  in this ``localization"   step then one is introducing spurious solutions.  The point is that, as we already discussed in sect.~\ref{sect:sol},
an equation such as $\Box U=-R$ is solved by \eq{Uhom},
where $U_{\rm hom}(x)$ is the general  solution of $\Box U_{\rm hom}=0$ and $G(x;x')$ is any a Green's function of the $\Box$ operator, and the definition of the nonlocal model is completed only once we have fully specified  the definition of $\iBox$, by specifying $U_{\rm hom}(x)$.  The specific choice of 
$U_{\rm hom}(x)$ can depend on the specific problem that we are studying, e.g. a different choice will appropriate for the study of a static solution, or for the study of FRW solutions, simply because the boundary conditions of these problems are different. In any case, in any given problem a choice must be performed, and  this fixes the  homogeneous solution. The solutions of the equations in the local formulation are also solutions of the original integro-differential equations only for this specific choice of  $U_{\rm hom}(x)$, and all other solutions are spurious.
Thus the solution of the homogeneous equation $\Box U=0$ cannot be interpreted as a free field, which is expanded in plane wave, and whose coefficients $a_{\vk}, a_{\vk}^{*}$ are then interpreted as  creation and annihilation operators in the corresponding quantum theory. Indeed, we have seen in sect.~\ref{sect:rmgg1}
that in the specific problem of a static spherically symmetric solution, the homogeneous solutions is fixed to the expression given in \eq{NewtU}. This is similar to what happens for the scalar metric perturbations $\Phi$ and $\Psi$, which are also fixed by the boundary conditions. In contrast,  for  the  tensor perturbations, on any given background we always have the freedom to add gravitational waves freely propagating to infinity, i.e. to add to the solution of \eq{hijsij} an arbitrary solution of the homogeneous equation $\Box h_{ij}^{\rm TT}=0$.

A similar choice must be made when we define what it means exactly to extract the transverse part
in \eq{modela2}. Indeed, the   solution of the homogeneous equation ${\cal D}^{\mu}_{\nu}S_{\mu}=0$, where ${\cal D}^{\mu}_{\nu}=(1/2) (\d^{\mu}_{\nu}\Box +\n^{\mu}\n_{\nu})$, is not a free radiative degree of freedom, but it is part of the definition of extraction of the transverse part. In this case the most obvious definition is to set this homogeneous solution to zero, corresponding to the fact that, if $S_{\mu\nu}=0$, there is no transverse part to extract, and $S^T_{\mu\nu}=0$, so we define the extraction of the transverse part so that, when $S_{\mu\nu}=0$, also $S_{\mu}=0$. 

Having realized that the fields $U$ and $S_{\mu}$ do not carry radiative degrees of freedom, it becomes clear that the content of \eq{loc1}, as far as radiative degrees of freedom are concerned, is the same as in GR, namely two massless graviton states with helicities $\pm 2$, which are now coupled also to extra non-radiative fields.  This can be checked explicitly by looking at the linearized version of the theory.
Linearizing \eq{modela2} over Minkowski space we get
\be\label{line1}
{\cal E}^{\mu\nu,\rho\sigma}\hrs
-\frac{2}{3}\, m^2 P^{\mu\nu}P^{\rho\sigma}
\hrs=-16\pi G\TMN\, ,
\ee
where 
\be
P^{\mu\nu}=\eMN-\frac{\paM\paN}{\Box}\, ,
\ee
and $\Box$ is now  the   flat-space d'Alembertian. Let us examine first the scalar sector.
In the  linearized limit we can put the theory in a local form in an even simpler way, namely  introducing two auxiliary scalar fields
$U=-\iBox R$ and $S=-\iBox U$. Then, writing again the metric in terms of the Bardeen variables and the energy-momentum tensor as in \eqs{T0i}{T0i},
the equations of the linearized theory in the scalar sector can be rewritten as~\cite{Maggiore:2014sia}
\bees
\n^2\[\Phi-(m^2/6) S\]&=&-4\pi G\rho\, ,\label{dof1}\\
\Phi-\Psi-(m^2/3) S&=& -8\pi G\sigma\, ,\label{dof2}\\
(\Box+m^2)U&=&-8\pi G (\rho-3P)\, ,\label{dof3}\\
\Box S&=&-U\, .
\ees
The equations for $U$ is just the linearization of $\Box U=-R$ and, as explained above, we must discard its homogeneous solution. Hence, it does not describe radiative degrees of freedom. The same holds for $S$, which plays the role that in the full nonlinear theory is played by the four-vector $S_{\mu}$, and is defined so that $S=0$ when $U=0$. 
The Bardeen variables $\Phi$ and $\Psi$ still satisfy Poisson equations, as in GR, and therefore remain non-radiative. This should be contrasted with what happens in massive gravity with a Fierz-Pauli mass term, where $\Phi$ becomes radiative and satisfies a massive Klein-Gordon equation
$(\Box-m^2)\Phi=0$~\cite{Deser:1966zzb,Alberte:2010it,Jaccard:2012ut}. Thus, in the nonlocal theory there is no radiative degree of freedom in the scalar sector, while the vector and tensor sector are obviously not affected by the presence of $U$ and $S$. In particular, the graviton remains massless, as we see from
\eqs{Seff}{Delta}. Indeed, the first term in \eq{Delta} describes the matter-matter interaction induced by a massless helicity-2 field, while the second term contributes to the  matter-matter interaction
$\tilde{T}_{\mu\nu}(-k)
\tilde{D}^{\mu\nu \rho\s}(k)\tilde{T}_{\rho\sigma}(k)$, with a term
\be\label{TT}
\frac{1}{6}\tilde{T}(-k)\[ \frac{1}{k^2}-\frac{1}{k^2-m^2}\]
\tilde{T}(k)\, .
\ee
These two terms are induced by the massless field $S$ and by the massive field $U$, respectively. If $U$ were a radiative field, its contribution would be ghost-like, and one should worry about vacuum decay in the quantum theory. However, we have seen that $U$ is not radiative, and at the quantum level there are no creation and annihilation operators associated to it, and no quantum vacuum instability.\footnote{Of course, such a field can induce instabilities at the classical level. However, while the quantum vacuum decay would be a disaster for the consistency of the theory, classical  instabilities must be examined on a case-by-case basis, and can in fact even be welcome. This is particularly true in a cosmological setting, where the emergence of a phase of accelerated expansion is in a sense a classical instability. Indeed, the result of \cite{Maggiore:2013mea} show that, at the background level, the cosmological evolution of this theory is perfectly viable. In \cite{Dirian:2014ara} we will examine the cosmological perturbations in these nonlocal models, and we will see again that they are perfectly viable, and in agreement with the observation.
Similarly, the results of the present paper showed that  no dangerous instability develops in static solutions, and that GR is smoothly recovered at $r\ll m^{-1}$.} 

Another way to understand this result, again discussed in \cite{Foffa:2013sma}, is to observe that, linearizing around flat space, $\gmn=\emn+\hmn$,
we have $R=R^{(1)}+{\cal O}(h^2)$, where
$R^{(1)}=\pam\pan (\hMN-\eMN h)$. Therefore in the linearized theory
\bees
U&=&-\iBox R^{(1)}=h-\frac{1}{\Box}\pam\pan\hMN\nn\\
&=&\(\emn-\frac{1}{\Box}\pam\pan\)\hMN\, .
\ees
Comparing with \eq{defshmn} we see that, in the linearized theory, $U$ is the same as the non-radiative field $s$. The quadratic Lagrangian corresponding to  the linearized equation (\ref{line1}) is
\be\label{Lquadr}
{\cal L}_2=\frac{1}{2}\hmn{\cal E}^{\mu\nu,\rho\sigma}\hrs
-\frac{1}{3}\, m^2\( P^{\mu\nu}\hmn\)^2\, ,
\ee
and we see that the term proportional to $m^2$ is just a mass term for $s=P^{\mu\nu}\hmn$. 
Writing the metric as in \eq{decomphmn}, instead of \eq{SEH2nl} we now obtain
\cite{Foffa:2013sma}
\be
S^{(2)}=\frac{1}{2}\int d^{4}x \,\[\hmn^{\rm TT}\Box (h^{\mu\nu })^{\rm TT}
-\frac{2}{3}\, s(\Box+m^2) s\]\, .
\ee
We see that the effect of the non-local term at the linearized level can be described as follows. In usual massless GR we have six physical (i.e. gauge-invariant) degrees of freedom, that can be described by the
five degrees of freedom of the transverse-traceless tensor $\hmn^{\rm TT}$, plus the scalar $s$. Out of them, only two are radiative, i.e. the helicity $\pm 2$ components of $\hmn^{\rm TT}$. The 
helicity-$(\pm 1)$ components of $\hmn^{\rm TT}$, the helicity-$0$ component of $\hmn^{\rm TT}$ and the scalar $s$ are all non-radiative, and all these six fields are massless. In the matter-matter interaction, the contribution from $s$ has the opposite sign, compared to the one coming from the helicity-$0$ component of $\hmn^{\rm TT}$, and the two cancel, while the helicity $\pm 1$ components of $\hmn^{\rm TT}$, are coupled to $\paM\Tmn$, which vanish, and therefore do not contribute. Thus, we only remain with the 
contribution from the  helicity-$(\pm 2)$ components.
In the nonlocal theory, the field $s$ remains non-radiative, but becomes massive. Therefore, the cancelation with the helicity-$0$ component of $\hmn^{\rm TT}$ is only approximate, and only holds for $|k^2|\gg m^2$. Thus, for $m\sim H_0$, well inside the horizon we recover GR and there is no vDVZ discontinuity, as indeed the computation of the previous sections showed explicitly. In contrast, at cosmological scales, there are departures from GR, but no new propagating degree of freedom.

It is also interesting to compare the above discussion with the results of refs.~\cite{Biswas:2011ar,Biswas:2013kla}, where the author performed a very general analysis of ghost-free modified gravitational actions, linearized over Minkowski space, by including the most general form factors depending on the 
$\Box$ operator, i.e. terms such as $\hmn a(\Box)\hMN$, 
$h_{\mu}^{\sigma}b(\Box)\pas\pan\hmn$, etc. The 
propagator can then be found in full generality, and one can impose conditions on the form factors $a(\Box), b(\Box)$, etc. 
such that no ghost-like pole appears, and furthermore the UV behavior is improved. 
The analysis in \cite{Biswas:2011ar,Biswas:2013kla} was mostly tuned toward the UV behavior, and therefore one is mostly concerned with positive powers of the $\Box$ operator. What we learn from our discussion in this section is that, when we apply this analysis to the IR, where non-local operators such as $\iBox$ become relevant, an apparent ghost-like pole in the propagator is not yet necessarily a sign of a trouble, since it could simply correspond to a non-propagating degree of freedom.

Observe also that, with our $(-,+,+,+)$ signature, the operator $(\Box+m^2 )$ that appears in the linearized equation of motion for $s$ (which is just \eq{dof3}, given that at the linearized level $s=U$) corresponds to a dispersion relation $k_0^2=-m^2+\vk^2$. Therefore, static solution do not decay at large distances with a Yukawa suppression $r^{-1} \exp\{-mr\}$, but are instead oscillatory, $r^{-1}\cos (mr)$, as indeed we found in sect.~\ref{sect:rmgg1}.

\section{Conclusions}\label{sect:concl}

The analytic and numerical results discussed in this paper show that, in the nonlocal theory
defined by \eq{modela2}, the linearized expansion is valid for all distances $r$ in the range $r_S\ll r\ll m^{-1}$, and in this region the corrections to GR are of the form $1+{\cal O}(m^2r^2)$. This is in sharp contrast with what typically happens in local theories of massive gravity, where the linear expansion breaks down below a Vainshtein radius $r_V$ which is parametrically larger than $r_S$, and which diverges as $m\ra 0$. In local massive gravity theories (whether in a Fierz-Pauli or dRGT form) this breakdown of linearity is  necessary for their observational viability, since these theories have a vDVZ discontinuity at large distance. Without such a breakdown of linearity, this discontinuity would persist down to the solar system scale, and then the theory would be ruled out. In contrast, the nonlocal theory (\ref{modela2}) has no vDVZ discontinuity, and it remains linear down to the near-horizon region. Therefore, all successes of GR at the solar system and lab scales are automatically recovered.
This is an important consistency check of the nonlocal theory which, together with its cosmological properties discussed in \cite{Maggiore:2013mea,Foffa:2013vma}, makes it a interesting candidate for a dynamical explanation of dark energy. 

Furthermore, we have determined the behavior of the solution in the region ($r\gg r_S$ and $mr$ generic) using a Newtonian expansion. \Eqs{NewtA}{NewtB} provide an analytic expression for the modifications of the static Newtonian forces at distances of order $m^{-1}$ in the nonlocal model that we have studied, and could have potential applications in the study of structure formation at large scales in such a model.

\vspace{5mm}

\noindent
{\bf Acknowledgments.}  We thank Claudia de~Rham for very useful comments.
The work of M.M. is supported by the Fonds National Suisse.
The  research of A.K. was implemented under the ``Aristeia" Action of the 
``Operational Programme Education and Lifelong Learning''
and is co-funded by the European 
Social Fund (ESF) and National Resources.  It is partially
supported by European Union's Seventh Framework Programme (FP7/2007-2013) under REA
grant agreement n. 329083.
%\newpage

\bibliographystyle{utphys}
\bibliography{myrefs_massive}

\end{document}